\newcommand{\shortversion}[1]{}
\newcommand{\longversion}[1]{#1}

\documentclass[a4paper,11pt]{article}

\shortversion{
\usepackage[a4paper,bindingoffset=0.2in,left=0.9in,right=0.97in,top=1in,bottom=1in,footskip=.25in]{geometry}
}

\usepackage{amssymb,amsmath,amsthm}
\usepackage[usenames,dvipsnames]{color}

\usepackage[comma,numbers,square,sort]{natbib} 
\usepackage{multirow}
\usepackage{mathtools}
\usepackage{nicefrac}

\longversion{\usepackage{fullpage}}

\usepackage{thmtools, thm-restate}

\DeclareMathOperator*{\argmax}{arg\,max}

\usepackage{makecell}

\usepackage{enumitem}

\usepackage[usenames,svgnames,dvipsnames]{xcolor}

\shortversion{

\usepackage{titlesec}

\titlespacing{\section}{0pt}{1pt}{1pt}
\titlespacing{\subsection}{0pt}{1pt}{1pt}
\titlespacing{\subsubsection}{0pt}{1pt}{1pt}
\titlespacing{\paragraph}{0pt}{1pt}{1pt}

\usepackage{titling}
\setlength{\droptitle}{-70pt}

\setlength{\textfloatsep}{0ex}
\setlength{\floatsep}{0ex}
\setlength{\intextsep}{1ex}

\AtBeginDocument{%
 \abovedisplayskip=0pt plus 0pt minus 0pt
 \abovedisplayshortskip=0pt plus 0pt
 \belowdisplayskip=0pt plus 3pt minus 0pt
 \belowdisplayshortskip=7pt plus 3pt minus 4pt
}
}

\usepackage{authblk}

\usepackage{url}

\usepackage{ifxetex}

\ifxetex
      \usepackage{euler}
      \usepackage{fontspec}
      \usepackage{xunicode}
      \defaultfontfeatures{Mapping=tex-text} 
      \newfontfamily\specialfont{Merge Light}
    \newfontfamily\fancyheading{Lorena}

    \setromanfont[SmallCapsFont={Overlock SC}]{Overlock}

\else
\usepackage{charter}
\fi

\usepackage{MnSymbol,wasysym}

\newcommand{\E}[1]{\mathbb{E}\left[#1\right]}

\newtheorem{observation}{\bf Observation}
\newtheorem{theorem}{\bf Theorem}
\newtheorem{lemma}{\bf Lemma}
\newtheorem{corollary}{\bf Corollary}
\newtheorem{definition}{\bf Definition}
\newtheorem{claim}{\bf Claim}

\newcommand{\nfrac}{\nicefrac}

\newcommand{\eps}{\varepsilon}
\renewcommand{\epsilon}{\eps}

\newcommand{\ignore}[1]{}

\title{Fishing out Winners from Vote Streams}

\author{\shortversion{\vspace{-15pt}}Arnab Bhattacharyya and Palash Dey}
\affil{\shortversion{\vspace{-10pt}}Indian Institute of Science, Bangalore}
\affil{\shortversion{\vspace{-10pt}}\texttt {\{arnabb,palash\}@csa.iisc.ernet.in}}

\date{}

\begin{document}

\maketitle

\shortversion{\vspace{-35pt}}
\begin{abstract}
We investigate the problem of winner determination from computational social choice theory in the data stream model. Specifically, we consider the task of summarizing an arbitrarily ordered stream of $n$ votes on $m$ candidates into a small space data structure so as to be able to obtain the winner determined by popular voting rules. As we show, finding the exact winner requires storing essentially all the votes. So, we focus on the problem of finding an {\em $\eps$-winner}, a candidate who could win by a change of at most $\eps$ fraction of the votes. We show non-trivial upper and lower bounds on the space complexity of $\eps$-winner determination for several voting rules, including $k$-approval, $k$-veto, scoring rules, approval, maximin, Bucklin, Copeland, and plurality with run off. 
\end{abstract}

\section{Introduction}

A common and natural way to aggregate preferences of agents is through an {\em election}.  In a typical election, we have a set of $m$ candidates and a set of $n$ voters, and each voter reports his ranking of the candidates in the form of a {\em vote}. A {\em voting rule} selects one candidate as the winner once all voters provide their votes. Determining the winner of an election is one of the most fundamental  problems in social choice theory. 

We consider elections held in an online setting where voters vote in arbitrary order, and we would like to find the winner at any point in time. A very natural scenario where this occurs is an election conducted over the Internet. For instance, websites often ask for rankings of restaurants in a city and would like to keep track of the ``best'' restaurant according to some fixed voting rule. Traditionally, social choice theory addresses settings where the number of candidates is much smaller than the number of voters. However, we now often have situations where both the candidate set and voter set are very large. For example, the votes may be the result of high-frequency measurements made by sensors in a network \citep{lesser2012distributed}, and a voting rule could be used to aggregate the measurements (as argued in \cite{caragiannis2011voting}). Also, in online participatory democracy systems, such as \citep{widescope, synapp}, the number of candidates can be as large as the number of voters. The na\"ive way to conduct an online election is to store all the vote counts in a database and recompute the winner whenever it is needed. The space complexity of this approach becomes infeasible if the number of candidates or the number of votes is too large. Can we do better? Is it possible to compress the votes into a short summary that still allows for efficient recovery of the winner?

This question can be naturally formulated in the {\em data stream model} \citep{henzinger1999computing, Alon1999137}. Votes are interpreted as items in a data stream, and the goal is to devise an algorithm with minimum space requirement to determine the election winner. In the simplest setting of the plurality voting rule, where each vote is simply an approval for a single candidate and the winner is the one who is approved by the most, our problem is closely related to the classic problem of finding heavy hitters \citep{charikar2004finding, cormode2005improved} in a stream. For other popular voting rules, such as {\em Borda, Bucklin} or {\em Condorcet consistent} voting rules, the questions become somewhat different.

Regardless of the voting rule, if the goal is to recover only the winner and the stream of votes is arbitrary, then it becomes essentially impossible to do anything better than the above-mentioned na\"ive solution (even when the algorithm is allowed to be randomized). Although we prove this formally, the reason should be intuitively clear: the winner may be winning by a very tiny margin thereby making every vote significant to the final outcome. We therefore consider a natural relaxation of the winner determination problem, where the algorithm is allowed to output any candidate who could have been the winner, according to the voting rule under consideration, by a change of at most $\eps n$ votes. We call such a candidate an {\em $\eps$-winner}; similar notions were introduced in \citep{xia2012computing, lee2014crowdsourcing}. Note that if the winner wins by a {\em margin of victory} \cite{xia2012computing} of more than $\eps n$, there is a unique $\eps$-winner.

In this work, we study streaming algorithms to solve the {\em \textsc{$(\eps,\delta)$-winner determination}} problem, i.e.~the task of determining, with probability at least $1-\delta$, an $\eps$-winner of any given vote stream according to popular voting rules. Our algorithms are necessarily randomized.

\subsection{Our Contributions}

We initiate the study of streaming algorithms for the \textsc{$(\epsilon, \delta)$--Winner Determination} problem with respect to various voting rules.  The results for the \textsc{$(\epsilon, \delta)$--Winner Determination} problem, when both $\eps$ and $\delta$ are positive, are summarized in \MakeUppercase table\nobreakspace \ref {tbl:summary}. (When $\eps$ or $\delta$ equals $0$, we prove that the space requirements are much larger.) 

We also exhibit algorithms, having space complexity nearly same as \MakeUppercase table\nobreakspace \ref {tbl:summary}, for the more general {\em sliding window} model, introduced by Datar et.~al.~in \cite{datar2002maintaining}. In this setting, for some parameter $N$, we want to find an $\eps$-winner with respect to the $N$ most recent votes in the stream, clearly a very well motivated scenario in online elections.

\begin{table}[t]
 \begin{center}
  \resizebox{\textwidth}{!}{
   \begin{tabular}{|c|c|c|}\hline
   
      \multirow{2}{*}{\textbf{Voting Rule}}	& \multicolumn{2}{c|}{\textbf{Space complexity}} \\\cline{2-3}
       & Upper bound & Lower bound \\\hline\hline
      
      &&\\[-10pt]
      
      Generalized plurality$^\ddag$ & \makecell{$O\big(\min \{ \frac{1}{\eps}\log m, m\log\frac{1}{\eps}\} + \log\log n\big)$ \\~[\MakeUppercase \bf Theorem\nobreakspace \ref {thm:kapp} and\nobreakspace  \ref {thm:gen_plu}]}
      & 
      \makecell{$\Omega(\frac{1}{\eps}\log\frac{1}{\eps} + \frac{1}{\sqrt{\eps}}\log m + \log\log n)$ if $m\ge \frac{1}{\eps}^\dagger$\\~[\MakeUppercase theorem\nobreakspace \ref {thm:eps_logm}]}
      
      \\\hline
      
      &&\\[-10pt]
      
      $k$-veto$^\star$ & \makecell{$O(\min \{ \frac{k}{\eps}\log m, m\log\frac{\log(m-k+1)}{\eps}\} + \log\log n)$\\~[\MakeUppercase theorem\nobreakspace \ref {thm:kapp}]} &
      \makecell{$\Omega(\frac{1}{\eps^\mu}\log\frac{1}{\eps} + \log m + \log\log n) \text{ if } m\ge \frac{1}{\eps}^\dagger$\\
      for every $\mu\in[0,1)$ [\MakeUppercase theorem\nobreakspace \ref {thm:veto_lb}]}
      \\\hline
      
      &&\\[-10pt]
      
      Plurality &  \makecell{$O\big(\min \{ \frac{1}{\eps}\log m, m\log\frac{1}{\eps}\} + \log\log n\big)$ \\~[\MakeUppercase \bf Theorem\nobreakspace \ref {thm:kapp} and\nobreakspace  \ref {thm:gen_plu}]}
      & \multirow{6}{*}{\makecell{\\ \\ \\ \\
      $
	\begin{cases}
	 \Omega(\frac{1}{\eps}\log\frac{1}{\eps} + \log m + \log\log n) & \text{ if } m\ge \frac{1}{\eps},\\
	 \Omega(m\log\frac{1}{\eps} + \log\log n)	& \text{ if } m\le \frac{1}{\eps},
	\end{cases}
      $\\ \\
      ~[\MakeUppercase \bf Theorem\nobreakspace \ref {thm:eps_eps} and\nobreakspace  \ref {thm:mlog_eps}]\\~[\MakeUppercase theorem\nobreakspace \ref {thm:loglogn}]\\~[\MakeUppercase observation\nobreakspace \ref {obs:trivial}]
      }}
      \\\cline{1-2}
      
      &&\\[-10pt]
      
      $k$-approval$^\star$ & \makecell{$O(\min \{ \frac{k}{\eps}\log m, m\log\frac{\log(k)}{\eps}\} + \log\log n)$\\~[\MakeUppercase theorem\nobreakspace \ref {thm:kapp}]} & \\\cline{1-2}
      
      &&\\[-10pt]
      
      Scoring rules & \makecell{$O(m(\log\log m + \log\frac{1}{\eps} ) + \log\log n)$\\~[\MakeUppercase theorem\nobreakspace \ref {thm:scr}]} & 
      \\\cline{1-2}
      
      &&\\[-10pt]
      
      Approval & \makecell{$O(m(\log\log m + \log\frac{1}{\eps} ) + \log\log n)$\\~[\MakeUppercase theorem\nobreakspace \ref {thm:app}]} &\\\cline{1-2}
      
      &&\\[-10pt]
      
      \makecell{Maximin, Bucklin, \\ Run off} & \makecell{$O(\min\{ m^2(\log\log m + \log\frac{1}{\eps}),$ \\ $\frac{1}{\eps^2}m\log^2 m \} + \log\log n)$ \\~[\MakeUppercase \bf Theorem\nobreakspace \ref {thm:store-samples} and\nobreakspace  \ref {thm:other}]} & \\\cline{1-2}
      
      &&\\[-10pt]
      
      Copeland & \makecell{ $O(\min\{m^2(\log\log m + \log\frac{1}{\eps}),$ \\ $\frac{1}{\eps^2}m\log^4 m \} + \log\log n)$ \\~[\MakeUppercase \bf Theorem\nobreakspace \ref {thm:store-samples} and\nobreakspace  \ref {thm:other}]} & \\\hline
   \end{tabular}
  }
  \caption{\small Space complexity for the \textsc{$(\epsilon, \delta)$-Winner Determination} problem for various voting rules. We do not show dependence on $\delta$ in the table for sake of clarity. $\star:$ The lower bound results for the $k$-approval and $k$-veto voting rules apply only for $k=O(m^\gamma)$, for every $\gamma\in[0,1)$. $\dagger$: For the case $m\le \frac{1}{\eps}$, the lower bound is same as that of other rules. $\ddag:$ Here, each voter has the choice to either approve or disapprove of one candidate and the candidate who has the maximum number of approvals minus disapprovals wins.}\label{tbl:summary}  
 \end{center}
\end{table}

\subsection{Related Work}

\subsubsection{Social Choice}
To the best of our knowledge, our work is the first to systematically study the approximate winner determination problem in the data stream model. A conceptually related work is that of Conitzer and Sandholm \cite{conitzer2005communication} who study the communication complexity of common voting rules. They consider $n$ parties each of whom knows only their own vote but, through a communication protocol, would like to compute the winner according to a specific voting rule. Observe that a streaming algorithm for exact winner determination using $s$ bits of memory space immediately\footnote{Each party can input its vote into the stream and then communicate the memory contents of the streaming algorithm to the next party.} implies a one-way communication protocol where each party transmits $s$ bits. However, it turns out that their results only imply weak lower bounds for the space complexity of streaming algorithms. Moreover, \cite{conitzer2005communication} does not study determination of $\eps$-winners. The communication complexity of voting rules was also highlighted by Caragiannis and Procaccia in \cite{caragiannis2011voting}.

In a recent work, we \cite{deysampling} studied the problem of determining election winners from a random sample of the vote distribution. Since we can randomly sample from a stream of votes using a small amount of extra storage, the bounds from \cite{deysampling} are also useful in the streaming context. In that work, the goal was to find the winner who was assumed to have a {\em margin of victory} \cite{xia2012computing} of at least $\eps$, but the same arguments also work for finding $\eps$-winners.

\subsubsection{Streaming}
The field of streaming algorithms has been the subject of intense research over the past two decades in both the algorithms and database communities. The theoretical foundations for the area were laid by \citep{henzinger1999computing, Alon1999137}. A stream is a sequence of data items $\sigma_1, \sigma_2, \dots, \sigma_n$, drawn from the universe $[m]$, such that on each {\em pass} through the stream, the items are read once in that order.  The frequency vector associated with the stream $f = (f_1, \cdots, f_m) \in \mathbb{Z}^m$ is defined as $f_j$ being the number of times $j$ occurs as an item in the stream. In this definition, the stream is {\em insertion-only}; more generally, in the {\em turnstile model}, items can both be inserted and deleted from the stream, in which case the frequency vector maintains the cumulative count of each element in $[m]$. General surveys of the area can be found in \cite{muthukrishnan2005data, nelson2012sketching}.

Algorithms for the insertion-only case were discovered before the formulation of the data streaming model. Consider the {\em point-query} problem: for a stream of $n$ items from a universe of size $m$ and a parameter $\eps > 0$, the goal is to output, for any item $j \in [m]$, an estimate $\hat{f}_j$ such that $|\hat{f}_j - f_j|\leq \eps n$. Misra and Gries \cite{misra82} gave\footnote{The algorithm can be viewed as a generalization of the Boyer-Moore \cite{boyer1991mjrty, fischer82finding} algorithm for $\eps = 1/2$.  It was also rediscovered 20 years later by \cite{demaine2002frequency, karp2003simple}.} an elegant but simple deterministic algorithm requiring only $O(\min\{m, {1}/{\eps}\} \cdot (\log m + \log n))$ space in bit complexity. Since to find an $\eps$-winner for the plurality voting rule, it's enough to solve the point query problem and output the $j$ with maximum $\hat{f}_j$, Misra-Gries automatically implies $O(\min\{m,{1}/{\eps}\} \cdot (\log m + \log n))$ space complexity for plurality. We use sampling to improve the dependence on $n$ and prove tightness in terms of $\eps$ and $n$. Our algorithms for many of the other voting rules are also based on the Misra-Gries algorithm. We note that in place of Misra-Gries, there are several other deterministic algorithms which could have been used, such as {\em Lossy Counting} \cite{manku2002approximate} and {\em Space Saving} \cite{metwally2005efficient}, but they would not change the asymptotic space complexity bounds.  A thorough overview of the point query, or frequency estimation, problem can be found in \cite{cormode2008finding}.

For the more general turnstile model, the point query problem for such streams is that of finding $\hat{f}_j$, for every $j$, such that $|\hat{f}_j - f_j| \leq \eps \|f\|_1$. The best result for this problem is due to Cormode and Muthukrishnan, the randomized {\em count-min sketch} \cite{cormode2005improved}, which has space complexity $O(\frac{1}{\eps} \log m \log n)$ in bits. The space bound was proved to be essentially tight by Jowhari et al.~in \cite{Jowhari}. In our context, the stream is a sequence of votes; so, our problems are mostly, just by definition, insertion-only. However, the count-min sketch becomes useful in our applications (i) if voters can issue retractions of their votes, and (ii) to maintain counts of random samples drawn from streams of unknown length.  

\subsection{Technical Overview}

\paragraph{Upper Bounds.}
The streaming algorithms that achieve the upper bounds shown in Table \ref{tbl:summary} are obtained through applying frequency estimation algorithms, such as Misra-Gries or count-min sketch, appropriately on a subsampled stream. The number of samples needed to obain $\eps$-winners for the various voting rules was previously analyzed in \cite{deysampling}.

\paragraph{Lower Bounds.}
Our main technical novelty is in the proofs of the lower bounds for the $(\eps, \delta)$-winner determination problem. Usually, in the ``heavy hitters'' problem in the algorithms literature, the task is {\em roughly} to determine the set of items with frequency above $\eps n$. Since there can be $1/\eps$ such items, a space lower bound of $\log {m \choose 1/\eps} = \Omega(\frac1\eps \log(\eps m))$ immediately follows for $m \gg 1/\eps$. In contrast, we wish to determine only one $\eps$-winner, so that just $\log m$ bits are needed to output the result. In order to obtain stronger lower bounds that depend on $\eps$, we need to resort to other techniques. Moreover, note that our lower bounds are in the insertion-only stream model, whereas previous lower bounds for frequency estimation problems are usually for the more general turnstile model.

We prove these bounds through new reductions from fundamental problems in communication complexity. To give a flavor of the reductions, let us sketch the proof for the plurality voting rule. Consider each additive term separately in the lower bound.

\begin{itemize}[topsep=0pt,leftmargin=12pt,itemsep=0pt]
\item
{$\boldsymbol{\log \log n}$}: Suppose Alice has a number $1 \leq a \leq n$ and Bob a number $1 \leq b \leq n$, and Bob wishes to know whether $a > b$ through a protocol where communication is one way from Alice to Bob. It is known \cite{smirnov88, MiltersenNSW98} that Alice is required to send $\Omega(\log n)$ bits to Bob. We can reduce this problem to finding a $1/3$-winner in a plurality election among two candidates by having Alice push $2^a$ approvals for candidate $1$ into the stream and Bob pushing $2^b$ approvals for candidate $2$; the $\Omega(\log \log n)$ lower bound  follows.
\item
$\boldsymbol{(1/\eps) \log(1/\eps)}$ \textbf{when} $\boldsymbol{m \ge 1/\eps}$: Consider the \textsc{Indexing} problem over an arbitrary alphabet: Alice has a vector $x \in [t]^m$ and Bob an index $i\in [m]$, and Bob wants to find $x_i$ through a one-way protocol from Alice to Bob. Erg\"un et al \cite{ergun2010periodicity}, extending \cite{MiltersenNSW98}'s proof for the case of $t=2$, show Alice needs to send $\Omega(m \log t)$ bits. For $t = m = 1/\sqrt{\eps}$, we reduce \textsc{Indexing} to $\eps$-winner determination for a plurality election. Let the candidate set be $[t]\times[m]$. Alice (given her input $x$) pushes $n/2$ votes into the stream with $\sqrt{\eps} n/2$ votes to each $(x_j, j)$ for all $j \in [m]$ and sends over the memory content of the streaming algorithm to Bob who (given his input $i$) pushes another $n/2$ votes into the stream with $\sqrt{\eps} n/2$ votes to each $(a, i)$ for all $a \in [t]$. Note that candidate $(x_i, i)$ is the unique $\sqrt{\eps}/4$-winner of this plurality election! Using \cite{ergun2010periodicity}'s lower bound $\Omega(1/\sqrt{\eps} \log(1/\eps))$ on the communication complexity of the \textsc{Indexing} problem yields our result.
\item
$\boldsymbol{m \log(1/\eps)}$ \textbf{when} $\boldsymbol{m \leq 1/\eps}$: Suppose Alice has a vector $a \in [t]^m$ and Bob a vector $b \in [t]^m$, and Bob wants to find\footnote{Assume the maximum is unique.} $i = \arg \max_j (a_j + b_j)$ through a one-way protocol. We show by reducing from the \textsc{Augmented Indexing} problem \cite{ergun2010periodicity, MiltersenNSW98} that Alice needs to send $\Omega(m \log t)$ bits to Bob. Suppose $t = 1/\eps$. Alice imagines her vector $a$ as being the vote count for a plurality election among $m$ candidates, streams in $a$ and runs the streaming algorithm for the problem, and passes the memory output to Bob who also streams in his vector $b$.  The maximum entry in $a+b$ corresponds to a candidate winning by margin at least $\eps^2 n$, hence yielding the $\Omega(m \log(1/\eps))$ lower bound.
\end{itemize}

\section{Preliminaries}

\subsection{Voting and Voting Rules}

Let $\mathcal{V}=\{v_1, \dots, v_n\}$ be the set of all \emph{voters} and $\mathcal{C}=\{c_1, \dots, c_m\}$ 
the set of all \emph{candidates}. If not mentioned otherwise, $\mathcal{V}$, $\mathcal{C}$, $n$ and $m$ denote set of voters, the set of candidates, the number of voters and the number of candidates respectively. Each voter $v_i$'s \textit{vote} is a  complete order $\succ_i$ over the candidate set $\mathcal{C}$. For example, for two candidates $a$ and $b$, $a \succ_i b$ means that the voter $v_i$ prefers $a$ to $b$. We denote the set of all complete orders over $\mathcal{C}$ by $\mathcal{L(C)}$. 
Hence, $\mathcal{L(C)}^n$ denotes the set of all $n$-voters' preference profiles $(\succ_1, \dots, \succ_n)$. 
A map $r:\uplus_{n,|\mathcal{C}|\in\mathbb{N}^+}\mathcal{L(C)}^n\longrightarrow 2^\mathcal{C}$
is called a \emph{voting rule}. Given a vote profile $\succ \in
\mathcal{L}(\mathcal{C})^n$, we call the candidates in $r(\succ)$ the {\em winners}. 
Given an election $\mathcal{E}=(\mathcal{V},\mathcal{C})$, we can construct a weighted graph $\mathcal{G}_\mathcal{E}$, called 
\textit{weighted majority graph}, from $\mathcal{E}$. The set of vertices in $\mathcal{G}_\mathcal{E}$ is the set of candidates in $\mathcal{E}$. For any two candidates $x$ and $y$, the weight on the edge $(x,y)$ is $D_\mathcal{E}(x,y) = N_\mathcal{E}(x,y) - N_\mathcal{E}(y,x)$, where $N_\mathcal{E}(x,y)$ (respectively $N_\mathcal{E}(y,x)$) is the number of voters who prefer $x$ to $y$ (respectively $y$ to $x$). A candidate $x$ is called the {\em Condorcet winner} in an election $\mathcal{E}$ if $D_\mathcal{E}(x,y) > 0$ for every other 
candidate $y \ne x$. A voting rule is called {\em Condorcet consistent} if it selects the Condorcet winner 
as the winner of the election whenever it exists. Some examples of common voting rules are:
\begin{itemize}[topsep=0pt,leftmargin=12pt,itemsep=0pt]

\item \textbf{Positional scoring rules:} A collection of $m$-dimensional vectors $\vec{s}_m=\left(\alpha_1,\alpha_2,\dots,\alpha_m\right)\in\mathbb{R}^m$  with $\alpha_1\ge\alpha_2\ge\dots\ge\alpha_m$ and $\alpha_1>\alpha_m$ for every $m\in \mathbb{N}$ naturally defines a voting rule -- a candidate gets score $\alpha_i$ from a vote if it is placed at the $i^{th}$ position. The 
 score of a candidate is the sum of the scores it receives from all the votes. 
 The winners are the candidate with maximum score. The vector $\alpha$ that is $1$ in the first $k$ coordinates and $0$ in other coordinates gives the {\em $k$-approval}  voting rule. The vector $\alpha$ that is $1$ in the last $k$ coordinates and $0$ in other coordinates is called {\em $k$-veto} voting rule. Observe that the score of a candidate in the $k$-approval (respectively $k$-veto) voting rule is the number of approvals (and respectively vetoes) that the candidate receives. $1$-approval is called the {\em plurality} voting rule, and $1$-veto is called the {\em veto} voting rule. The score vector $(m-1, m-2, \dots, 1, 0)$ gives the {\em Borda} rule.
 
\item \textbf{Generalized plurality:} In generalized plurality voting, each voter approves or disapprove one candidate. The score of a candidate is the number of approvals it receives minus number of disapprovals it receives. The candidates with highest score are the winners. We introduce this rule and consider it to be interesting particularly in an online setting where every voter either likes or dislikes an item; hence each vote is either an approval for a candidate or a disapproval for a candidate.
 
\item \textbf{Approval:} In approval voting, each voter approves a subset of candidates. The winners are the candidates which are approved by the maximum number of voters. 
 
\item \textbf{Maximin:} The maximin score of a candidate $x$ is $\min_{y\ne x} D_{\mathcal{E}}(x,y)$. The winners are the candidates with maximum maximin score. 
 
\item \textbf{Copeland:} The Copeland score of a candidate $x$ is $|\{y\ne x:D_{\mathcal{E}}(x,y)>0\}|$. The winners are the candidates with maximum Copeland score.
 
\item \textbf{Bucklin:} A candidate $x$'s Bucklin score is the minimum number $\ell$ such that more than half 
 of the voters rank $x$ in their first $\ell$ positions. The winners are the candidates with lowest Bucklin score.
 
\item \textbf{Plurality with runoff:} The top two candidates according to
 plurality score are selected first. The pairwise winner of these two
 candidates is selected as the winner of the election. This rule is
 often called the {\em runoff} voting rule.
\end{itemize}

Among the above, only the maximin and Copeland 
rules are Condorcet consistent.

\subsection{Model of Input Data}

In the basic model, the input data is an insertion only stream of elements from some universe $\mathcal{U}$. We note that, in the context of voting in an online scenario, the natural model of input data is the insertion only streaming model over the universe of all possible votes $\mathcal{L(C)}$. The basic model can be generalized to the more sophisticated {\em sliding window model} where the only {\em active} items are the last $n$ items, for some parameter $n$. In this work, we focus on winner determination algorithms for insertion only stream of votes in both basic and sliding window models. The basic input model can also be generalized to another input model, called {\em turnstile model}, where the input data is a sequence from $\mathcal{U}\times\{1,-1\}$; every element in the stream corresponds to either a unit increment or a unit decrement of frequency of some element from $\mathcal{U}$. We will use the turnstile streaming model (over some different universe) only to design efficient winner determination algorithms for the insertion only stream of votes. We note that, the algorithms for the streaming data can make only one pass over the input data. These one pass algorithms are also called {\em streaming algorithms}.

\subsection{Communication Complexity}

We will use lower bounds on {\em communication complexity} of certain functions to prove space complexity lower bounds for our problems. Communication complexity of a function measures the number of bits that need to be exchanged between two players to compute a function whose input is split among those two players~\citep{yao1979some}. In a more restrictive {\it one-way communication model}, the first player sends only one message to the second player and the second player outputs the result. A protocol is a method that the players follow to compute certain functions of their input. Also the protocols can be randomized; in that case, the protocol needs to output correctly with probability at least $1-\delta$, for some parameter $\delta\in[0,1]$ (the probability is taken over the random coin tosses of the protocol). The randomized one-way communication complexity of a function $f$ with error $\delta$ is denoted by $\mathcal{R}_\delta^{1-way}(f)$. Classically the first player is named Alice and the second player is named Bob and we also follow the same convention here. \citep{Kushilevitz} is a standard reference for communication complexity.

\longversion{
\subsection{Chernoff Bound}

We will use the following concentration inequality:

\begin{theorem}\label{thm:chernoff}
Let $X_1, \dots, X_\ell$ be a sequence of $\ell$ independent
random variables in $[0,1]$ (not necessarily identical). Let $S = \sum_i X_i$ and
let $\mu = \E{S}$. Then, for any $0 \leq \delta \leq 1$: 
$$\Pr[|S-\mu| \geq \delta \ell] < 2 \exp(-2\ell \delta^2)$$
and 
$$\Pr[|S - \mu| \geq \delta \mu] < 2\exp(-\delta^2\mu/3)$$
The first inequality is called an additive bound and the second
multiplicative. 
\end{theorem}
}
 
\subsection{Problem Definition}

The basic winner determination problem is defined as follows.
 
\begin{definition}\textsc{(Winner Determination)}\\
 Given a voting profile $\succ$ over a set of candidates $\mathcal{C}$ and a voting rule $r$, determine\longversion{ the winners} $r(\mathcal{\succ})$.
\end{definition}

We show a strong space complexity lower bound for the \textsc{Winner Determination} problem for the plurality voting rule in \MakeUppercase theorem\nobreakspace \ref {thm:exact-lwb}. To overcome this theoretical bottleneck, we focus on determining {\em approximate winner} of an election. Below we define the notion of $\eps$-approximate winner which we also call {\em $\eps$-winner}.

\begin{definition}\textsc{($\epsilon$-winner)}\\
 Given an $n$-voter voting profile $\succ$ over a set of candidates $\mathcal{C}$ and a voting rule $r$, a candidate $w$ is called an $\epsilon$--winner if $w$ can be made winner by changing at most $\epsilon n$ votes in $\succ$.
\end{definition}

Notice that there always exist an $\epsilon$-winner in every election since a winner is also an $\epsilon$-winner. We show that finding even an $\eps$-winner deterministically requires large space when the number of votes is large [see \MakeUppercase theorem\nobreakspace \ref {thm:logn}]. However, we design space efficient randomized algorithms which outputs an $\eps$-winner of an election with probability at least $1-\delta$. The problem that we study here is called \textsc{$(\epsilon, \delta)$-Winner Determination} problem and is defined as follows.

\begin{definition}\textsc{($(\epsilon, \delta)$-Winner Determination)}\\
 Given a voting profile $\succ$ over a set of candidates $\mathcal{C}$ and a voting rule $r$, determine an $\epsilon$--winner with probability at least $1-\delta$.  (The probability is taken over the internal coin tosses of the algorithm.)
\end{definition}

\section{Upper Bounds}

In this section, we present the algorithms for the $(\epsilon, \delta)$-Winner Determination problem for various voting rules.\shortversion{ Because of space constraints, we move proofs of some of the results (which are marked $\star$) to the appendix.} Before embarking on specific algorithms, we first prove a few supporting results that will be used crucially in our algorithms later. We begin with the following space efficient algorithm for picking an item uniformly at random from a universe of size $n$ below.

\begin{restatable}{observation}{ObsSamplingUB}\shortversion{[$\star$]}\label{lem:sampling_ub}
 There is an algorithm for choosing an item with probability $\frac{1}{n}$ that uses $O(\log\log n)$ bits of memory and uses fair coin as its only source of randomness. 
\end{restatable}

\longversion{
\begin{proof}
 First let us assume, for simplicity, that $n$ is a power of $2$. We toss a fair coin $\log_2 n$ many times and choose the item, say $x$, only if the coin comes head all the times. Hence the probability that the item $x$ gets chosen is $\frac{1}{n}$. We need $O(\log\log n)$ space to toss the fair coin $\log_2 n$ times (to keep track of the number of times we have tossed the coin so far). If $n$ is not a power of $2$ then, toss the fair coin $\lceil \log_2 n \rceil$ many times and we choose the item $x$ only if the coin comes head in all the tosses conditioned on some event $E$. The event $E$ contains exactly $n$ outcomes including the all heads outcome.
\end{proof}
}

We remark that \MakeUppercase observation\nobreakspace \ref {lem:sampling_ub} is tight in terms of space complexity. We state the claim formally below, as it may be interesting in its own right.
\begin{restatable}{proposition}{PropSamplingLB}\shortversion{[$\star$]}\label{lem:sampling_lb}
 Any algorithm that chooses an item from a set of size $n$ with probability $p$, for $0< p \le \frac{1}{n}$, using a fair coin as its only source of randomness, must use $\Omega(\log\log n)$ bits of memory. 
\end{restatable}

\longversion{
\begin{proof}
 The algorithm tosses the fair coin some number of times (the number of times it tosses the coin may also depend on the outcome of the previous tosses) and finally picks an item from the set. Consider a run $\mathcal{R}$ of the algorithm where it chooses the item, say $x$, with {\em smallest number of coin tosses}; say it tosses the coin $t$ many times in this run $\mathcal{R}$. This means that in any other run of the algorithm where the item $x$ is chosen, the algorithm must toss the coin at least $t$ number of times. Let the outcome of the coin tosses in $\mathcal{R}$ be $r_1, \cdots, r_t$. Let $s_i$ be the memory content of the algorithm immediately after it tosses the coin $i^{th}$ time, for $i\in [t]$, in the run $\mathcal{R}$. First notice that if $t < \log_2 n$, then the probability with which the item $x$ is chosen is more than $\frac{1}{n}$, which would be a contradiction. Hence, $t \ge \log_2 n$. Now we claim that all the $s_i$'s must be different. Indeed otherwise, let us assume $s_i = s_j$ for some $i<j$. Then the algorithm chooses the item $x$ after tossing the coin $t-(j-i)$ (which is strictly less than $t$) many times when the outcome of the coin tosses are $r_1, \cdots, r_i, r_{j+1}, \cdots, r_t$. This contradicts the assumption that the run $\mathcal{R}$ we started with chooses the item $x$ with smallest number of coin tosses.
\end{proof}
}

An essential ingredient in our algorithms is calculating the approximate frequencies of all the elements in a universe in an input data stream. The following result (due to~\citep{misra82}) provides a space efficient algorithm for that job.

\begin{restatable}{theorem}{ThmMisra}\shortversion{[$\star$]}\label{thm:misra}
 Given an insertion only stream of length $n$ over a universe of size $m$, there is a deterministic one pass algorithm to find the frequencies of all the items in the stream within an additive approximation of $\eps n$ using $O\left(\min\{\frac{1}{\eps}\left(\log m + \log n \right), m\log n\}\right)$ bits of memory, for every $\eps>0$. 
\end{restatable}

\longversion{
\begin{proof}
 The $O\left(\frac{1}{\eps}\left(\log m + \log n \right)\right)$ space algorithm is due to~\citep{misra82}. On the other hand, notice that with space $O\left(m\log n \right)$, we can exactly count the frequency of every element, {\em even in the turnstile model of stream}, by simply keeping an array of length $m$ (indexed by ids of the elements from the universe) each entry of which is capable of storing integers up to $n$.
\end{proof}
}

We now describe streaming algorithms for the \textsc{$(\epsilon, \delta)$--Winner Determination} problem for various voting rules. The general idea is to sample certain number of votes uniformly at random from the stream of votes using the algorithm of \MakeUppercase observation\nobreakspace \ref {lem:sampling_ub} and generate another stream of {\em elements} over some different universe. The number of votes sampled and the universe of the stream generated depend on the specific voting rule we are considering. After that, we approximately calculate the frequencies of the elements in the generated stream using \MakeUppercase theorem\nobreakspace \ref {thm:misra}. For simplicity, we assume that the number of votes in known in advance up to some constant factor (only to be able to apply \MakeUppercase observation\nobreakspace \ref {lem:sampling_ub}). We will see in Section\nobreakspace \ref {sec:unknown} how to get rid of this assumption, without affecting space complexity of any of the algorithms much. We begin with the $k$-approval and $k$-veto voting rules below.

\begin{restatable}{theorem}{KappUbThm}\shortversion{[$\star$]}\label{thm:kapp}
 Assume that the number of votes is known to be within $[c_1n, c_2n]$ for some constants $c_1$ and $c_2$ in advance. Then there is a one pass algorithm for the \textsc{$(\epsilon, \delta)$--Winner Determination} problem for the $k$-approval voting rule that uses $O\left( \min\{\frac{k}{\eps}\left( \log m + \log\frac{1}{\eps} + \log\log\frac{1}{\delta} \right), m\left(\log\frac{\log(k+1)}{\eps} + \log\log\frac{1}{\delta} \right)\} + \log\log n \right)$ bits of memory and for the $k$-veto voting rule that uses $O\big( \min\{\frac{k}{\eps}\left( \log m + \log\frac{1}{\eps} + \log\log\frac{1}{\delta} \right), m \big( \log\frac{\log(m-k+1)}{\eps} + \log\log\frac{1}{\delta} \big)\} + \log\log n \big)$ bits of memory.
\end{restatable}

\longversion{
\begin{proof}
 Let us first consider the case of the $k$-approval voting rule. We pick the current vote in the stream with probability $p$ (the value of $p$ will be decided later) independent of other votes. Suppose we sample $\ell$ many votes; let $\mathcal{S} = \{ v_i : i\in[\ell] \}$ be the set of votes sampled. From the set of sampled votes $\mathcal{S}$, we generate a stream $\mathcal{T}$ over the universe $\mathcal{C}$ as follows. For $i\in [\ell]$, let the vote $v_i$ be $c_1\succ c_2\succ \cdots \succ c_m$. From the vote $v_i$, we add $k$ candidates $c_1, \cdots, c_k$ in the stream $\mathcal{T}$. We know that there is a $\ell = O(\frac{\log(k+1)}{\eps^2} \log \frac{1}{\delta})$ (and thus a corresponding $p=\Omega(\frac{1}{n})$) which ensures that for every candidate $x\in \mathcal{C}$, $|\frac{s(x)}{n} - \frac{\hat{s}(x)}{\ell}| < \frac{\eps}{3}$ with probability at least $1 - \frac{\delta}{2}$~\citep{deysampling}, where $s(\cdot)$ and $\hat{s}(\cdot)$ are the scores of the candidates in the input stream of votes and in $\mathcal{S}$ respectively. Now we count $\hat{s}(x)$ for every candidate $x\in \mathcal{C}$ within an additive approximation of $\frac{\eps \ell}{3}$ and the result follows from \MakeUppercase theorem\nobreakspace \ref {thm:misra} (notice that the length of the stream $\mathcal{T}$ is $k\ell$). 
 
 For the $k$-veto voting rule, we approximately calculate the number of vetoes that every candidate gets using the same technique as above. However, for the $k$-veto voting rule, the corresponding bound for $\ell$ is $O(\frac{\log(m-k+1)}{\eps^2} \log \frac{1}{\delta})$ which implies the result.
\end{proof}
}

By similar techniques, we have the following algorithm for the generalized plurality rule.

\begin{restatable}{theorem}{ThmGenPluUB}\shortversion{[$\star$]}\label{thm:gen_plu}
 Assume that the number of votes is known to be within $[c_1n, c_2n]$ for any constants $c_1$ and $c_2$ in advance. Then there is a one pass algorithm for the \textsc{$(\epsilon, \delta)$--Winner Determination} problem for the generalized plurality voting rule that uses $O\left( \frac{1}{\eps}\left(\log m + \log \frac{1}{\eps} + \log\log \frac{1}{\delta} \right) + \log\log n\right)$ bits of memory.
\end{restatable}

\longversion{
\begin{proof}
 We sample $\ell=O(\frac{1}{\eps^2}\log\frac{1}{\delta})$ many votes uniformly at random from the input stream of votes using the technique used in the proof of \MakeUppercase theorem\nobreakspace \ref {thm:kapp}. For every candidate, we count both the number of approvals and disapprovals that it gets within an additive approximation of $\frac{\eps\ell}{10}$ which is enough to get an $\eps$-winner. Now the space complexity follows form \MakeUppercase theorem\nobreakspace \ref {thm:misra}.
\end{proof}
}

We generalize \MakeUppercase theorem\nobreakspace \ref {thm:kapp} to the class of scoring rules next. We need the following result in the subsequent proof which is due to~\citep{deysampling}. 

\begin{lemma}\label{lem:scr}
 Let $\alpha = (\alpha_1, \cdots, \alpha_m)$ be an arbitrary score vector and $w$ the winner of an $\alpha$--election $\mathcal{E}$. Let $x$ be any candidate which is not a $\eps$--winner. Then, $ s(w) - s(x) \ge \alpha_1 \eps n $.
\end{lemma}

With \MakeUppercase lemma\nobreakspace \ref {lem:scr} at hand, we now present the algorithm for the scoring rules.

\begin{theorem}\label{thm:scr}
 Assume that the number of votes is known to be within $[c_1n, c_2n]$ for any constants $c_1$ and $c_2$ in advance. Let $\alpha = (\alpha_1, \cdots, \alpha_m)$ be a score vector such that $\alpha_i\ge 0$ for every $i\in[m]$. Then there is a one pass algorithm for the \textsc{$(\epsilon, \delta)$--Winner Determination} problem for the $\alpha$-scoring rule that uses $O\left( \frac{\sum_{i=1}^m \alpha_i}{\alpha_1}\left( \log\log m + \log\frac{1}{\eps} + \log\log\frac{1}{\delta} \right) + \log\log n \right)$, which is $O\left( m\left( \log\log m + \log\frac{1}{\eps} + \log\log\frac{1}{\delta} \right) + \log\log n \right)$, bits of memory.
\end{theorem}

\shortversion{\vspace{-15pt}}\begin{proof}
 Let $\alpha = (\alpha_1, \cdots, \alpha_m)$ be an arbitrary score vector with $\alpha_i\ge 0$ for every $i\in[m]$. We define $\alpha_i^\prime = \frac{\alpha_i}{\sum_{i=j}^m \alpha_j}$ (which is in $[0,1]$), for every $i\in[m]$. Since scoring rules remain same even if we multiply every $\alpha_i$ with any positive constant $\lambda$, the score vectors $\alpha$ and $\alpha^\prime$ correspond to same voting rule. We pick the current vote in the stream with probability $p$ (the value of $p$ will be decided later) independent of other votes. Suppose we sample $\ell$ many votes; let $\mathcal{S} = \{ v_i : i\in[\ell] \}$ be the set of votes sampled. For $i\in [\ell]$, let the vote $v_i$ be $c_1\succ c_2\succ \cdots \succ c_m$. We pick the candidate $c_i$ from the vote $v_i$ with probability $\alpha_i^\prime$ and define it to be $a_i$. We compute the frequencies of the candidates in the stream $\bar{\mathcal{S}} = \{ a_i : i\in [\ell] \}$ within an additive factor of $\eps^\prime n$, where $\eps^\prime = \frac{\eps}{3}$. For every candidate $x\in \mathcal{C}$, let $s(x)$ be the $\alpha^\prime$--score of the candidate $x$ in the input stream of votes and $\hat{s}(x)$ be $\frac{n}{\ell}$ times the $\alpha^\prime$--score of the candidate $x$ in the sampled votes $\mathcal{S}$. We know that there exists an $\ell = O(\frac{1}{\eps^2} \log \frac{m}{\delta})$ (and thus a corresponding $p=\Omega(\frac{1}{n})$) which ensures that, for every candidate $x\in \mathcal{C}$, $|s(x) - \hat{s}(x)| < \alpha_1^\prime \eps^\prime n$ with probability at least $1 - \frac{\delta}{2}$~\citep{deysampling}. Let $\bar{s}(x)$ be $\frac{n}{\ell}$ times the frequency of the candidate $x\in \mathcal{C}$ in the stream $\bar{\mathcal{S}}$. We now prove the following claim from which the result follows immediately.
 \begin{claim}\label{clm:scr}
  \[ \Pr[ \forall x\in \mathcal{C}, |\bar{s}(x) - \hat{s}(x)| \le \alpha_1^\prime \eps^\prime n ] \ge 1 - \frac{\delta}{2} \]
 \end{claim}
 \shortversion{\vspace{-15pt}}\begin{proof}
  For every candidate $x\in \mathcal{C}$ and every $i\in [\ell]$, we define a random variable $X_i(x)$ to be $1$ if $a_i = x$ and $0$ otherwise. Then, $\bar{s}(x) = \frac{n}{\ell} \sum_{i\in [\ell]} X_i(x)$. We have, $\E{\bar{s}(x)} = \hat{s}(x)$. Now using Chernoff bound\longversion{ from \MakeUppercase theorem\nobreakspace \ref {thm:chernoff}}, we have the following:
  \longversion{
  \begin{align*}
   \Pr[ |\bar{s}(x) - \hat{s}(x)| > \alpha_1^\prime \eps^\prime n ] &= \Pr[ |\frac{n}{\ell} \sum_{i\in [\ell]} X_i(x) - \hat{s}(x)| > \alpha_1^\prime \eps^\prime n ]\\
   &= \Pr[ |\sum_{i\in [\ell]} \frac{X_i(x)}{\alpha_1^\prime} - \frac{ \ell\hat{s}(x)}{\alpha_1^\prime n}| > \eps^\prime \ell ]\\
   &\le 2 \exp\{ -\frac{\eps^2 \alpha_1^\prime n \ell}{3 \hat{s}(x)} \}\\
   &\le 2 \exp\{ -\frac{\eps^2 \ell}{3} \}
  \end{align*}
  The fourth inequality follows from the fact that $\hat{s}(x) \le \alpha_1^\prime n$ for every candidate $x\in \mathcal{C}$. Now we use the union bound to get the following.
  }
  \shortversion{
  \[\Pr[ |\bar{s}(x) - \hat{s}(x)| > \alpha_1^\prime \eps^\prime n ] = \Pr[ |\frac{n}{\ell} \sum_{i\in [\ell]} X_i(x) - \hat{s}(x)| > \alpha_1^\prime \eps^\prime n ] \le 2 \exp\{ -\frac{\eps^2 \alpha_1^\prime n \ell}{3 \hat{s}(x)} \} \le 2 \exp\{ -\frac{\eps^2 \ell}{3}\}\]
  The third inequality follows from the fact that $\hat{s}(x) \le \alpha_1^\prime n$ for every candidate $x\in \mathcal{C}$. Now we use the union bound to get the following.
  }
  \[ \Pr[ \forall x\in \mathcal{C}, |\bar{s}(x) - \hat{s}(x)| \le \alpha_1^\prime \eps^\prime n ] \ge 1 - \sum_{x\in \mathcal{C}}2 \exp\{ -\frac{\eps^2 \ell}{3}\} \ge 1 - \frac{\delta}{2} \]
 The second inequality follows from an appropriate choice of $\ell = O(\frac{1}{\eps^2} \log \frac{m}{\delta})$.
 \end{proof}
 We estimate the frequency of every candidate in $\bar{\mathcal{S}}$ within an additive approximation ratio of $\alpha_1^\prime\eps\ell$ and output the candidate $w$ with maximum estimated frequency as the winner of the election. The candidate $w$ is an $\eps$-- winner (follows from \MakeUppercase lemma\nobreakspace \ref {lem:scr}) with probability at least $1 - \delta$ (follows from \MakeUppercase claim\nobreakspace \ref {clm:scr}). The space complexity of this algorithm follows from \MakeUppercase theorem\nobreakspace \ref {thm:misra} (since $\frac{1}{\alpha_1^\prime} = \frac{\sum_{i=1}^m \alpha_i}{\alpha_1} \le \frac{m\alpha_1}{\alpha_1} = m$) and \MakeUppercase observation\nobreakspace \ref {lem:sampling_ub}.
\end{proof}

We present next the streaming algorithm for the approval voting rule. It is again obtained by running a frequency estimation algorithm on samples from a stream.

\begin{restatable}{theorem}{ThmApprovalUB}\shortversion{[$\star$]}\label{thm:app}
 Assume that the number of votes is known to be within $[c_1n, c_2n]$ in advance, for some constants $c_1$ and $c_2$. Then there is a one pass algorithm for the \textsc{$(\epsilon, \delta)$--Winner Determination} problem for the approval voting rule that uses $O\left( m\left( \log\log m + \log\frac{1}{\eps} + \log\log\frac{1}{\delta} \right) + \log\log n \right)$ bits of memory. 
\end{restatable}

\longversion{
\begin{proof}
 We sample $\ell$ many votes using the algorithm described in \MakeUppercase observation\nobreakspace \ref {lem:sampling_ub} and technique described in the proof of \MakeUppercase theorem\nobreakspace \ref {thm:scr}. The total number of approvals in those sampled votes is at most $m\ell$ and we estimate the number of approvals that every candidate receives within an additive approximation of $\frac{\eps\ell}{2}$. The result now follows from the upper bound on $\ell$~\citep{deysampling} and \MakeUppercase theorem\nobreakspace \ref {thm:misra}.
\end{proof}
}

Now we move on to maximin, Copeland, Bucklin, and plurality with run off voting rules. We provide two algorithms for these voting rules, which trade off between the number of candidates $m$ and the approximation factor $\eps$. The algorithm in \MakeUppercase theorem\nobreakspace \ref {thm:store-samples} below, which has better space complexity when $\frac{1}{\eps}$ is small compared to $m$, simply stores all the sampled votes.

\begin{restatable}{theorem}{ThmStoreUB}\shortversion{[$\star$]}\label{thm:store-samples}
 Assume that the number of votes is known to be within $[c_1n, c_2n]$ in advance, for some constants $c_1$ and $c_2$. Then there is a one pass algorithm for the \textsc{$(\epsilon, \delta)$--Winner Determination} problem for the maximin, Bucklin, and plurality with run off voting rules that use $O\left( \frac{m\log^2 m \log\frac{1}{\delta}}{\eps^2} + \log\log n \right)$ bits of memory and for the Copeland voting rule that uses $O\left( \frac{m\log^4 m \log\frac{1}{\delta}}{\eps^2}  + \log\log n \right)$ bits of memory. 
\end{restatable}

\longversion{
\begin{proof}
 We sample $\ell$ many votes from the input stream of votes uniformly at random and simply store all of them. Notice that we can store a vote using space $O(m\log m)$. The result now follows from the upper bound on $\ell$~\citep{deysampling} and \MakeUppercase observation\nobreakspace \ref {lem:sampling_ub}.
\end{proof}
}

Next we consider the case when $\frac{1}{\eps}$ is large compared to $m$.

\begin{theorem}\label{thm:other}
 Assume that the number of votes is known to be within $[c_1n, c_2n]$ in advance, for some constants $c_1$ and $c_2$. Then there is a one pass algorithm for the \textsc{$(\epsilon, \delta)$--Winner Determination} problem for the maximin, Copeland, Bucklin, and  plurality with runoff voting rules that uses $O\left( m^2\left( \log\log m + \log\frac{1}{\eps} + \log\log\frac{1}{\delta} \right) + \log\log n \right)$ bits of memory.
\end{theorem}

\shortversion{\vspace{-15pt}}\begin{proof}
 For each voting rule mentioned in the statement, we sample $\ell$ many votes $\mathcal{S} = \{ v_i : i\in[\ell] \}$ uniformly at random from the input stream of votes using the algorithm used in \MakeUppercase observation\nobreakspace \ref {lem:sampling_ub} and the technique used in the proof of \MakeUppercase theorem\nobreakspace \ref {thm:scr}. From $\mathcal{S}$, we generate another stream $\bar{\mathcal{S}}$ of elements belonging to a different universe $\mathcal{U}$ (which depends on the voting rule under consideration). Finally, we calculate the frequencies of the elements of $\bar{\mathcal{S}}$, using \MakeUppercase theorem\nobreakspace \ref {thm:misra}, within an additive approximation of $\frac{\eps \ell}{2}$ for maximin, Bucklin, and plurality with runoff voting rules and $\frac{\eps \ell}{2\log m}$ for the Copeland voting rule. The difference of approximation factor is due to~\citep{deysampling}. We know that $\ell=O\left(\frac{\log \frac{m}{\delta}}{\eps^2}\right)$ for maximin, Bucklin, and plurality with run off voting rules and $\ell=O\left(\frac{\log^3 \frac{m}{\delta}}{\eps^2}\right)$ for the Copeland voting rule~\citep{deysampling}. This bounds on $\ell$ prove the result once we describe $\bar{\mathcal{S}}$ and $\mathcal{U}$. Below, we describe the stream $\bar{\mathcal{S}}$ and the universe $\mathcal{U}$ for individual voting rules. Let the vote $v_i$ be $c_1\succ c_2\succ \cdots \succ c_m$.
 \begin{itemize}[topsep=2pt,leftmargin=15pt,itemsep=0pt]
  \item {\bf maximin, Copeland:} $\mathcal{U} = \mathcal{C} \times \mathcal{C}$. From the vote $v_i$, we put $(c_j, c_k)$ in $\bar{\mathcal{S}}$ for every $j< k$.
  \item {\bf Bucklin:} $\mathcal{U} = \mathcal{C} \times [m]$. From the vote $v_i$, we put $(c_j, k)$ in $\bar{\mathcal{S}}$ for every $j\le k$.
  \item {\bf plurality with runoff:} $\mathcal{U} = \mathcal{C} \times \mathcal{C}$. From the vote $v_i$, we put $(c_j, c_k)$ in $\bar{\mathcal{S}}$ for every $j< k$ and $(c_1, c_1)$. In the plurality with runoff voting rule, we need to estimate the plurality score of every candidate which we do by estimating the frequencies of the elements of the $(x, x)$ in $\bar{\mathcal{S}}$. We also need to estimate $D_\mathcal{E}(x, y)$ for every candidate $x, y\in \mathcal{C}$ which we do by estimating the frequencies of the elements of the form $(x,y)$.\qedhere
 \end{itemize}
\end{proof}

\subsection{Unknown stream length}\label{sec:unknown}

Now we consider the case when the number of voters is not known beforehand. The idea is to use reservoir sampling (\citep{Vitter}) along with approximate counting (\citep{Morris,flajolet1985approximate}) to pick an element from the stream {\em almost uniformly} at random. The following result shows that we can do so in a space efficient manner.

\begin{theorem}(Theorem 7 of \citep{GronemeierS09})\label{thm:sampling-unbd}
 Given an insertion only stream of length $n$ ($n$ is not known to the algorithm beforehand) over a universe of size $m$, there is a randomized one pass algorithm that outputs, with probability at least $1-\delta$, the element at a random position $X\in [n]$ such that, for every $i\in [n], |\Pr\{X=i\}-\frac{1}{n}|\le \frac{\eps}{n}$ using $O(\log\frac{1}{\delta}+\log\frac{1}{\eps}+\log\log n + \log m)$ bits of memory, for every $\eps \in (0,1]$ and $\delta>0$.
\end{theorem}

Recall that \MakeUppercase theorem\nobreakspace \ref {thm:misra} only works for insertion only streams. However, as the stream progresses, the element chosen by \MakeUppercase theorem\nobreakspace \ref {thm:sampling-unbd} changes; so, we cannot invoke Misra-Gries to do frequency estimation on a set of samples given by \MakeUppercase theorem\nobreakspace \ref {thm:sampling-unbd}. For streams with both insertions and deletions, we have the following result which is due to count-min sketch~\citep{cormode2005improved}.

\begin{theorem}\label{thm:count-min}
 Given a turnstile stream of length $n$ over a universe of size $m$, there is a randomized one pass algorithm to find the frequencies of the items in the stream within an additive approximation of $\eps n$ with probability at least $1-\delta$ using $O\left(\frac{\log m}{\eps}\log(\frac{1}{\delta})\left(\log m + \log n \right)\right)$ bits of memory, for every $\eps>0$ and $\delta>0$.
\end{theorem}

From \MakeUppercase \bf Theorem\nobreakspace \ref {thm:sampling-unbd} and\nobreakspace  \ref {thm:count-min} and from the proofs of \MakeUppercase \bf Theorem\nobreakspace \ref {thm:misra},   \ref {thm:gen_plu} to\nobreakspace  \ref {thm:app}  and\nobreakspace  \ref {thm:other}, we get the following.

\begin{restatable}{corollary}{CorNUnknown}\shortversion{[$\star$]}\label{cor:n_unknown}
 Assume that the number of votes $n$ is not known beforehand. Then there is a one pass algorithm for the \textsc{$(\epsilon, \delta)$--Winner Determination} problem for $k$-approval, $k$-veto, generalized plurality, approval, maximin, Copeland, Bucklin, and plurality with run off voting rules that uses $\log m\log\frac{1}{\delta}$ times more space than the corresponding algorithms when $n$ is known beforehand upto a constant factor. 
\end{restatable}

\longversion{
\begin{proof}
 We use reservoir sampling with approximate counting from \MakeUppercase theorem\nobreakspace \ref {thm:sampling-unbd}. The resulting stream that we generate have both positive and negative updates (since in reservoir sampling, we sometimes replace an item we previously sampled). Now we approximately estimate the frequency of every item in the generated stream using \MakeUppercase theorem\nobreakspace \ref {thm:count-min}.
\end{proof}
}

Again from \MakeUppercase \bf Theorem\nobreakspace \ref {thm:store-samples} and\nobreakspace  \ref {thm:sampling-unbd}, we get the following result which provides a better space upper bound than \MakeUppercase corollary\nobreakspace \ref {cor:n_unknown} when the number of candidates $m$ is large.

\begin{corollary}\label{cor:unkown_store}
 Assume that the number of votes $n$ is not known beforehand. Then there is a one pass algorithm for the \textsc{$(\epsilon, \delta)$--Winner Determination} problem for the maximin, Bucklin, and plurality with run off voting rules that use $O\left( \frac{m\log^2 m\log\frac{1}{\delta}}{\eps^2} + \log\log n \right)$ bits of memory and for the Copeland voting rule that uses $O\left( \frac{m\log^4 m\log\frac{1}{\delta}}{\eps^2}  + \log\log n \right)$ bits of memory.
\end{corollary}

\subsection{Sliding Window Model} 

Suppose we want to compute an $\eps$-winner of the last $n$ many votes in an infinite stream of votes for various voting rules. The following result shows that there is an algorithm, with space complexity same as \MakeUppercase theorem\nobreakspace \ref {thm:sampling-unbd}, to sample a vote from the last $n$ votes in a stream.

\begin{theorem}(\citep{Braverman})\label{thm:slide_sample}
 Given an insertion only stream over a universe of size $m$, there is a randomized one pass algorithm that outputs, with probability at least $1-\delta$, the element at a random position $X$ from last $n$ positions such that, for every $i\in [n], |\Pr\{X=i\}-\frac{1}{n}|\le \frac{\eps}{n}$ using $O(\log\frac{1}{\delta}+\log\frac{1}{\eps}+\log\log n + \log m)$ bits of memory, for every $\eps \in (0,1]$ and $\delta>0$.
\end{theorem}

\MakeUppercase theorem\nobreakspace \ref {thm:slide_sample} immediately provides results same as \MakeUppercase \bf Corollary\nobreakspace \ref {cor:n_unknown} and\nobreakspace  \ref {cor:unkown_store}, where $n$ is the window size.

\section{Lower Bounds}\label{subsec:lwb}

In this section, we prove space complexity lower bounds for the \textsc{$(\epsilon, \delta)$--Winner Determination} problem for various voting rules. We reduce certain communication problems to the \textsc{$(\epsilon, \delta)$--Winner Determination} problem for proving space complexity lower bounds. Let us first introduce those communication problems with necessary results.

\subsection{Communication Complexity}

\begin{definition}(\textsc{Augmented-indexing}$_{m,t}$)\\
 Let $t$ and $m$ be positive integers. Alice is given a string $x = (x_1, \cdots, x_t)\in [m]^t$. Bob is given an integer $i\in [t]$ and $(x_1, \cdots, x_{i-1})$. Bob has to output $x_i$.
\end{definition}

The following communication complexity lower bound result is due to~\citep{ergun2010periodicity} by a simple extension of the arguments of Bar-Yossef et al \cite{bar2002information}.

\begin{lemma}\label{lem:aug}
$\mathcal{R}_\delta^{1-way}(\textsc{Augmented-indexing}_{m,t}) = \Omega((1-\delta)t \log m)$ for any $\delta < 1 - \frac{3}{2m}$.
\end{lemma}

Also, we recall the multi-party version of the set-disjointness problem.
 \begin{definition}(\textsc{Disj}$_{m,t}^{promise}$)\\
  We have $t$ sets $X_1, \cdots, X_t$ each a subset of $[m]$. We have $t$ players and player $i$ is holding the set $X_i$. We are also given the promise that either $X_i \cap X_j = \emptyset$ for every $i\ne j$ or there exist an element $y\in [m]$ such that $y\in X_i$ for every $i\in [t]$ and $(X_i\setminus\{y\}) \cap (X_j\setminus\{y\}) = \emptyset$ for every $i\ne j$. The output \textsc{Disj}$_{m,t}^{promise}$($X_1, \cdots, X_t$) is $1$ if $X_i \cap X_j = \emptyset$ for every $i\ne j$ and $0$ else.
 \end{definition}
  
 \begin{lemma}[Proved in \cite{bar2002information,chakrabarti2003near}.] \label{lem:disj_t}
  $\mathcal{R}_\delta^{1-way}(\textsc{Disj}_{m,t}^{promise}) = \Omega(\frac{m}{t})$, for any $\delta\in[0,1)$ and $t$.
 \end{lemma}

The following communication problem is very useful for us.
\begin{definition}(\textsc{Max-sum}$_{m,t}$)\\\label{def:maxsum}
 Alice is given a string $x=(x_1, x_2, \cdots, x_t)\in [m]^t$ of length $t$ over universe $[m]$. Bob is given another string $y=(y_1, y_2, \cdots, y_t)\in [m]^t$ of length $t$ over the same universe $[m]$. The strings $x$ and $y$ is such that the index $i$ that maximizes $x_i+y_i$ is unique. Bob has to output the index $i\in[t]$ which satisfies $x_i+y_i = \max_{j\in[t]}\{x_j+y_j\}$.
\end{definition}

We establish the following one way communication complexity lower bound for the \textsc{Max-sum}$_{m,t}$ problem by reducing it from the \textsc{Augmented-indexing}$_{2,t\log m}$ problem.

\begin{restatable}{lemma}{LemMaxSumLB}\shortversion{[$\star$]}\label{mel:maxsum}
 $\mathcal{R}_\delta^{1-way}(\textsc{Max-sum}_{m,t}) = \Omega(t\log m)$, for every $\delta < \frac{1}{4}$. 
\end{restatable}

\longversion{
\begin{proof}
We reduce the \textsc{Augmented-indexing}$_{2,t\log m}$ problem to \textsc{Max-sum}$_{8m,t+1}$ problem thereby proving the result. Let the inputs to Alice and Bob in the \textsc{Augmented-indexing}$_{2,t\log m}$ instance be $(a_1, a_2, \cdots, a_{t\log m})\in \{0,1\}^{t\log m}$ and $(a_1, \cdots, a_{i-1})$ respectively. The idea is to construct a corresponding instance of the \textsc{Max-sum}$_{8m,t+1}$ problem that outputs $t+1$ if and only if $a_i=0$. We achieve this as follows. Alice starts execution of the \textsc{Max-sum}$_{8m,t+1}$ protocol using the vector $x=(x_1, x_2, \cdots, x_{t+1})\in [8m]^{t+1}$ which is defined as follows: the binary representation of $x_j$ is $\left(0,0,a_{\left(j-1\right)\log m + 1}, a_{\left(j-1\right)\log m + 2}, a_{\left(j-1\right)\log m + 3}, \cdots, a_{j\log m}, 0\right)_2$, for every $j\in [t]$, and $x_{t+1}$ is $0$. Bob participates in the \textsc{Max-sum}$_{8m,t+1}$ protocol with the vector $y=(y_1, y_2, \cdots, y_{t+1})\in [8m]^{t+1}$ which is defined as follows. Let us define $\lambda = \lceil \frac{i}{\log m} \rceil$. We define $y_j=0$, for every $j\notin \{\lambda, t+1\}$. The binary representation of $y_\lambda$ is $(1,0, a_{(\lambda-1)\log m+1}, a_{(\lambda-1)\log m+2}, \cdots, a_{i-1}, 1, 0, 0, \cdots, 0, 0, 1)_2$. Let us define an integer $T$ whose binary representation is $(0,0, a_{(\lambda-1)\log m+1}, a_{(\lambda-1)\log m+2}, \cdots, a_{i-1}, 0, 1, 1, \cdots, 1)_2$. We define $y_{t+1}$ to be $T+y_\lambda$. First notice that the output of the \textsc{Max-sum}$_{8m,t+1}$ instance is either $\lambda$ or $t+1$, by the construction of $y$.  Now observe that if $a_i=1$ then, $x_\lambda>T$ and thus the output of the \textsc{Max-sum}$_{8m,t+1}$ instance should be $\lambda$. On the other hand, if $a_i=0$ then, $x_\lambda<T$ and thus the output of the \textsc{Max-sum}$_{8m,t+1}$ instance should be $t+1$.
\end{proof}
}

Finally, we also consider the \textsc{Greater-than} problem.
\begin{definition}(\textsc{Greater-than}$_{n}$)\\
 Alice is given an integer $x\in [n]$ and Bob is given an integer $y\in [n], y\ne x$. Bob has to output $1$ if $x>y$ and $0$ otherwise.
\end{definition}

The following result is due to \citep{smirnov88, MiltersenNSW98}. We provide a simple proof of it that seems to be missing\footnote{A similar proof appears in \cite{kremer1999randomized} but theirs gives a weaker lower bound.} in the literature.

\begin{restatable}{lemma}{LemGT}\shortversion{[$\star$]}\label{lem:gt}
$\mathcal{R}_\delta^{1-way}(\textsc{Greater-than}_{n}) = \Omega(\log n)$, for every $\delta < \frac{1}{4}$. 
\end{restatable}

\longversion{
\begin{proof}
 We reduce the \textsc{Augmented-indexing}$_{2,\lceil\log n\rceil + 1}$ problem to the \textsc{Greater-than}$_{n}$ problem thereby proving the result. Alice runs the \textsc{Greater-than}$_{n}$ protocol with its input number whose representation in binary is $a=(x_1x_2\cdots x_{\lceil\log n\rceil}1)_2$. Bob participates in the \textsc{Greater-than}$_{n}$ protocol with its input number whose representation in binary is $b=(x_1x_2\cdots x_{i-1}1\underbrace{0 \cdots 0}_{(\lceil\log n\rceil-i+1)~ 0's})_2$. Now $x_i=1$ if and only if $a>b.$
\end{proof}
}

\subsection{Reductions}

\subsubsection{The cases $\eps = 0$ and $\delta=0$}
We begin with the problem where we have to find the winner (i.e., 0-winner) for a plurality election. Notice that, we can find the winner by exactly computing the plurality score of every candidate. This requires $O(m \log n)$ bits of memory. We prove below that, when $n$ is much larger than $m$, this space complexity is {\em almost} optimal even if we are allowed to use randomization, by reducing it from the \textsc{Max-sum}$_{n,m}$ problem. This strengthens a similar result proved in Karp et al. \cite{karp2003simple} only for deterministic algorithms.

\begin{restatable}{theorem}{ThmExactLB}\shortversion{[$\star$]}\label{thm:exact-lwb}
 Any one pass \textsc{$(0, \delta)$--Winner Determination} algorithm for the plurality and generalized plurality election must use $\Omega(m\log(n/m))$ bits of memory, for any $\delta\in[0,\frac{1}{4})$. 
\end{restatable}

\longversion{
\begin{proof}
 We prove the result for \textsc{$(0, \delta)$--Winner Determination} problem for the plurality election. This gives the result for the generalized plurality election since every plurality election is also a generalized plurality election. Consider the \textsc{Max-sum}$_{n,m}$ problem where Alice is given a string $x=(x_1, \cdots, x_m)\in[n]^m$ and Bob is given another string $y=(y_1, \cdots, y_m)\in[n]^m$. The candidate set of our election is $[m]$. The votes would be such that the only winner will be the candidate $i$ such that $i\in \argmax_{j\in[m]}\{x_j+y_j\}$. Moreover, the winner would be known to Bob, thereby proving the result. Thus Bob can output $x_i$ correctly whenever our \textsc{$(0, \delta)$--Winner Determination} algorithm outputs correctly. Alice generates $x_j$ many plurality votes for the candidate $j$, for every $j\in[m]$. Alice now sends the memory content to Bob. Bob resumes the run of the algorithm by generating $y_j$ many plurality votes for the candidate $j$, for every $j\in[m]$. The plurality score of candidate $j$ is $(x_j+y_j)$ and thus the plurality winner will be a candidate $i$ such that $i\in \argmax_{j\in[m]}\{x_j+y_j\}$. Notice that the total number of votes is at most $2mn$. The result now follows from \MakeUppercase lemma\nobreakspace \ref {mel:maxsum}.
\end{proof}
}

For the case when $m$ and $n$ are comparable, the following result is stronger. We prove this by exhibiting a reduction from the \textsc{Disj}$_{m,3}^{promise}$ problem. 

\begin{restatable}{theorem}{ThmExactLBmn}\shortversion{[$\star$]}\label{thm:exact-lwb-mn}
  Any one pass \textsc{$(0, \delta)$--Winner Determination} algorithm for the plurality and generalized plurality election must use $\Omega(\min\{m,n\})$ bits of memory, for any $\delta\in[0,1)$. 
\end{restatable}

\longversion{
\begin{proof}
  Suppose we have a one pass \textsc{$(0, \delta)$--Winner Determination} algorithm for the plurality election that uses $s$ bits of memory. We will demonstrate a one-way three party protocol to compute \textsc{Disj}$_{m,3}^{promise}$ function using $2s$ bits of communication thus proving the result. We have the candidate set $[m+1]$. The protocol is as follows.
  
  Player $1$ starts running the one pass \textsc{$(0, \delta)$--Winner Determination} algorithm on the input $X_1\cup \{m+1\}$. Once player $1$ is done reading all its input, it sends its memory content to player $2$. This needs at most $s$ bits of communication. Player $2$ resumes the run of the algorithm with input $X_2\cup \{m+1\}$ and sends its memory content to player $3$. Again this needs at most $s$ bits of communication. Player $3$ resumes the run of the algorithm on input $X_3$ and output $1$ if and only if the winner is $m+1$ and $0$ else. Notice that, if the $X_i \cap X_j = \emptyset$ for every $i\ne j$ then, the only winner of the votes $(X_1, m+1, X_2, m+1, X_3)$ is the candidate $m+1$ with a plurality score of two. On the other hand, if there exist an element $y\in [m]$ such that $y\in X_i$ for every $i\in [t]$ and $(X_i\setminus\{y\}) \cap (X_j\setminus\{y\}) = \emptyset$ for every $i\ne j$ then, the only winner of the votes $(X_1, m+1, X_2, m+1, X_3)$ is the candidate $y$ with a plurality score of three.
  
  The number of candidates in the election above is $m+1$ and the number of votes $n$ is $|X_1|+|X_2|+|X_3|+2(m+1) = \Theta(m).$ This gives a space complexity lower bound of $\Omega(\min\{m,n\}).$
\end{proof}
}

\MakeUppercase \bf Theorem\nobreakspace \ref {thm:exact-lwb} and\nobreakspace  \ref {thm:exact-lwb-mn} give space complexity lower bounds for the case $\eps = 0$. Next, we consider the other extreme case: deterministically find an $\eps$-winner, corresponding to $\delta=0$.

\begin{restatable}{theorem}{ThmLogn}\shortversion{[$\star$]}\label{thm:logn}
 Assume $\eps < \frac{1}{5}$. Then any one pass \textsc{$(\eps, 0)$--Winner Determination} algorithm for the plurality election must use $\Omega(\log n)$ bits of memory, even if the number of voters is known up to a factor of $2$ and the number of candidates is only $2$. The same applies for generalized plurality, scoring rules, maximin, Copeland, Bucklin, and plurality with run off voting rules.
\end{restatable}

\longversion{
\begin{proof}
 For the sake of contradiction, we assume that the number of possible memory contents of the algorithm is $o(n)$, since otherwise the algorithm uses $\Omega(\log n)$ space and we have nothing to prove. Our candidate set is $\{0,1\}$.  We will generate two vote streams, say $R_1$ and $R_2$, in such a way that the final state of the algorithm would be same; however $\eps$--winner would be different for the two streams thus providing the contradiction we are looking for. 
 
 Let $s_0$ be the starting state of the algorithm. Consider the stream of votes for $1$ and let the algorithm repeats its state for the first time after reading $i$ many $1$ votes. Let the state of the algorithm after reading $i^{th}$ $1$ vote be same as the state the algorithm was after it read $j^{th}$ $1$ vote. Let us call $\mu = i-j$. Clearly $\mu = o(n)$. Then there exist $\delta_1, \delta_2 = o(n)$ such that the state the algorithm will be after reading $\frac{n}{4}-\delta_1$ many votes for $1$ is same as the state it will be after reading $\frac{3n}{4}+\delta_2$ many votes for $1$. Let $R_1$ be the stream of $\frac{n}{4}-\delta_1$ many votes for $1$ followed by $\frac{n}{2}$ many votes for $0$. Let $R_2$ be the stream of $\frac{3n}{4}+\delta_2$ many votes for $1$ followed by $\frac{n}{2}$ many votes for $0$. By construction the output of the algorithm is same for both the streams $R_1$ and $R_2$. However, candidate $1$ is only $\eps$-winner in $R_1$ and candidate $0$ is only $\eps$-winner in $R_2$.

For elections with two candidates, scoring rules, maximin, Copeland, Bucklin, and plurality with run off voting rules are same as the plurality voting rule.
\end{proof}
}

\subsubsection{Lower Bounds for Approximate and Randomized algorithms}

Now we move on and show space complexity lower bounds for general \textsc{$(\eps, \delta)$--Winner Determination} problem for various voting rules. The observation below immediately follows from the fact that the algorithm has to output a candidate as an $\eps$-winner.

\begin{observation}\label{obs:trivial}
 Every \textsc{$(\eps, \delta)$--Winner Determination} algorithm, for all the voting rules considered in this paper, needs $\Omega(\log m)$ bits of memory.
\end{observation}

We show next a space complexity lower bound of $\Omega(\frac{1}{\eps}\log \frac{1}{\eps})$ bits for the \textsc{$(\eps, \delta)$--Winner Determination} problem for various voting rules.

\begin{theorem}\label{thm:eps_eps}
 Suppose the number of candidates $m$ is at least $\frac{1}{\eps}$. Any one pass \textsc{$(\eps, \delta)$--Winner Determination} algorithm for approval, $k$-approval, for $k=O(m^\lambda)$ for every $\lambda\in [0,1)$, generalized plurality, Borda, maximin, Copeland, and plurality with run off elections must use $\Omega((1-\delta)\frac{1}{\eps}\log \frac{1}{\eps})$ bits of memory, even when the number of votes are exactly known beforehand, for every $1-\delta > \frac{3\eps}{2}$.
\end{theorem}

\shortversion{\vspace{-15pt}}\begin{proof}
 We will show that, when $m\ge\frac{1}{\eps}$, we need $\Omega(\frac{1}{\sqrt{\eps}}\log\frac{1}{\eps})$ bits of memory for solving the \textsc{$(\frac{\sqrt{\eps}}{8}, \delta)$--Winner Determination} problem, thereby proving the result. Consider the \textsc{Augmented-indexing}$_{\nfrac{1}{\sqrt{\eps}}, \nfrac{1}{\sqrt{\eps}}}$ problem where Alice is given a string $x=(x_1, x_2, \cdots, x_{\nfrac{1}{\sqrt{\eps}}})\in [\nfrac{1}{\sqrt{\eps}}]^{\nfrac{1}{\sqrt{\eps}}}$ and Bob is given an integer $i\in [\nfrac{1}{\sqrt{\eps}}]$ and $(x_1, \cdots, x_{i-1})$. The candidate set of the election, that we generate, is $[\nfrac{1}{\sqrt{\eps}}]\times[\nfrac{1}{\sqrt{\eps}}]$. The overview of the technique is as follows: Alice generates a stream of votes and runs the algorithm, then sends the memory content to Bob, and Bob resumes the run of the algorithm with another stream of votes (both the streams of votes depend on the voting rule under consideration) in such a way that the only $\frac{\sqrt{\eps}}{8}$--winner will be the candidate $(x_i, i)$. Thus Bob can output $x_i$ correctly if and only if the \textsc{$(\nfrac{\sqrt{\eps}}{8}, \delta)$--Winner Determination} algorithm outputs correctly. Now the result follows from \MakeUppercase lemma\nobreakspace \ref {lem:aug}. The elections for specific voting rules are as follows. Let $n$ be the number of votes. 
 \begin{itemize}[topsep=2pt,leftmargin=15pt]
  
  \item {\bf $k$-approval for $k=O(m^\lambda)$ for every $\lambda\in [0,1)$, approval, and generalized plurality:} It is enough to prove the result for the $k$-approval voting rule for $k=O(m^\lambda)$ for every $\lambda\in [0,1)$, since every $k$-approval election is also an approval election. For $k=1$, we get the result for the plurality voting rule and thus for the generalized plurality voting rule, since every plurality election is also a generalized plurality election.
  \begin{itemize}[topsep=2pt,leftmargin=12pt]
   \item {\bf Case 1:} $k\le \sqrt{m}$: 
   Alice generates a stream of $\frac{n}{2}$ votes in such a way that the $k$-approval score of every candidate in $\{(x_j,j) : j\in [\nfrac{1}{\sqrt{\eps}}]\}$ is at least $\lfloor{k\sqrt{\eps} n}/{2}\rfloor$ and the $k$-approval score of any other candidate is $0$. Alice now sends the memory content of the algorithm to Bob. Bob resumes the run of the algorithm by generating another stream of ${n}/{2}$ votes in such a way that the $k$-approval score of every candidate in $\{(j,i) : j\in [\nfrac{1}{\sqrt{\eps}}]\}$ is at least $\lfloor{k\sqrt{\eps} n}/{2}\rfloor$ and the $k$-approval score of any other candidate is $0$. The score of the candidate $(x_i, i)$ is at least $\lfloor k\sqrt{\eps} n\rfloor$ where as the score of every other candidate is at most $\lceil{k\sqrt{\eps} n}/{2}\rceil$. Hence the only $\nfrac{\sqrt{\eps}}{8}$--winner is $(x_i, i)$.
   \item {\bf Case 2:} $k > \sqrt{m}$ and $k=O(m^\lambda)$ for any $\lambda\in [0.5,1)$:
   Alice generates a stream of $\frac{n}{2}$ votes in such a way that the $k$-approval score of every candidate in $\{(x_j,j) : j\in [\frac{1}{\sqrt{\eps}}]\}$ is at least $\frac{n}{2}$ and the $k$-approval score of any other candidate is at most $\lceil(k-\frac{1}{\sqrt{\eps}}){n}/\left({\frac{2}{\sqrt{\eps}}(\frac{1}{\sqrt{\eps}}-1)}\right)\rceil$, which is at most $\frac{n}{2} - \frac{\sqrt{\eps}}{2}n$ for sufficiently small constant $\eps$ (depending on $\lambda$). Alice now sends the memory content of the algorithm to Bob. Bob resumes the run of the algorithm by generating another stream of $\frac{n}{2}$ votes in such a way that the $k$-approval score of every candidate in $\{(j,i) : j\in [\frac{1}{\sqrt{\eps}}]\}$ is at least $\frac{n}{2}$ and the $k$-approval score of any other candidate is $\lceil(k-\frac{1}{\sqrt{\eps}})\frac{n}{\frac{2}{\sqrt{\eps}}(\frac{1}{\sqrt{\eps}}-1)}\rceil$. In this case also the only $\frac{\sqrt{\eps}}{8}$--winner is $(x_i, i)$.
  \end{itemize}  
  
  \item {\bf Borda, Bucklin:} Alice generates a stream of $\frac{n}{2}$ votes where the candidates in $\{(x_\ell,\ell), \ell\in [\nfrac{1}{\sqrt{\eps}}]\}$ are uniformly distributed in top $\nfrac{1}{\sqrt{\eps}}$ positions of the votes and the rest of the candidates are uniformly distributed in bottom $\nfrac{1}{\eps}-\nfrac{1}{\sqrt{\eps}}$ positions of the votes. Alice now sends the memory content to Bob and Bob resumes the run of the algorithm by generating another stream of ${n}/{2}$ votes where the candidates in $\{(\ell,i), \ell\in [\nfrac{1}{\sqrt{\eps}}]\}$ are uniformly distributed in top $\nfrac{1}{\sqrt{\eps}}$ positions of the votes and the rest of the candidates are uniformly distributed in bottom $\nfrac{1}{\eps}-\nfrac{1}{\sqrt{\eps}}$ positions of the votes. The Borda score of the candidate $(x_i, i)$ is $(\nfrac{1}{\eps}-\nfrac{1}{2\sqrt{\eps}})n$ whereas the Borda score of every other candidate is at most $(\nfrac{1}{2\eps}-\nfrac{1}{4\sqrt{\eps}})n$. Hence, the only $\nfrac{\sqrt{\eps}}{8}$--winner for the Borda voting rule is $(x_i, i)$, since each vote change can reduce or increase the Borda score of any candidate by at most ${1}/{\eps}$. 
  
  The candidate $(x_i, i)$ is ranked within top $\nfrac{2}{3\sqrt{\eps}}$ positions in $\nfrac{2n}{3}$ many votes, whereas any other candidate is ranked within top $\nfrac{2}{3\sqrt{\eps}}$ positions in at most $\nfrac{n}{3}$ many votes. Hence the only $\nfrac{\sqrt{\eps}}{8}$--winner for the Bucklin voting rule is $(x_i, i)$.
  
  \item {\bf Any Condorcet consistent voting rule, Plurality with runoff:} Let us define $X = \{(x_\ell, \ell): \ell\in[\nfrac{1}{\sqrt{\eps}}]\}$, $Y = [\nfrac{1}{\sqrt{\eps}}]\times[\nfrac{1}{\sqrt{\eps}}] \setminus X$. Suppose $\overrightarrow{X}$ and $\overrightarrow{Y}$ are arbitrary but fixed ordering of the candidates in $X$ and $Y$ respectively. For every $\ell \in [\nfrac{1}{\sqrt{\eps}}]$, Alice generates ${\sqrt{\eps} n}/{4}$ votes of the form $(x_\ell, \ell) \succ \overrightarrow{X\setminus \{(x_\ell, \ell)\}} \succ \overrightarrow{Y}$ and another ${\sqrt{\eps} n}/{4}$ votes of the form $\overleftarrow{X\setminus (x_\ell, \ell)} \succ (x_\ell, \ell) \succ \overrightarrow{Y}$, where $\overleftarrow{X}$ is the reverse order of $\overrightarrow{X}$. Alice now sends the memory content to Bob. Let us define $A = \{(\ell, i): \ell\in[\nfrac{1}{\sqrt{\eps}}]\}$ and $B = [\nfrac{1}{\sqrt{\eps}}]\times[\nfrac{1}{\sqrt{\eps}}] \setminus A$. Suppose $\overrightarrow{A}$ and $\overrightarrow{B}$ are arbitrary but fixed ordering of $A$ and $B$ respectively. Bob resumes the run of the algorithm by generating another $\nfrac{\sqrt{\eps} n}{4}$ votes of the form $(\ell, i) \succ \overrightarrow{A\setminus (\ell, i)} \succ \overrightarrow{B}$ and another $\nfrac{\sqrt{\eps} n}{4}$ votes of the form $\overleftarrow{A\setminus (\ell, i)} \succ (\ell, i) \succ \overrightarrow{B}$ for every $\ell \in [\nfrac{1}{\sqrt{\eps}}]$, where $\overleftarrow{A}$ is the reverse order of $\overrightarrow{A}$. The candidate $(x_i, i)$ defeats every other candidate in pairwise election by a margin of at least $\frac{n}{4}$. Also the plurality score of the candidate $(x_i, i)$ is more than the plurality score of every other candidate by at least $\sqrt{\eps} n$. Hence the only $\nfrac{\sqrt{\eps}}{8}$--winner is $(x_i, i)$.\qedhere  
 \end{itemize} 
\end{proof}

We can prove a space lower bound of $\Omega(m\eps\log\frac{1}{\eps})$ for one pass \textsc{$(\eps, \delta)$--Winner Determination} algorithms for Borda, Bucklin, Copeland, and maximin voting rules by reducing it from \textsc{Augmented-indexing}$_{\nfrac{1}{\eps}, m}$ in the proof of \MakeUppercase theorem\nobreakspace \ref {thm:eps_eps}. We summarize this observation below.

\begin{corollary}
 Suppose the number of candidates $m$ is at least $\frac{1}{\eps}$. Any one pass \textsc{$(\eps, \delta)$--Winner Determination} algorithm for Borda, maximin, Copeland, and plurality with run off elections must use $\Omega((1-\delta)m\log \frac{1}{\eps})$ bits of memory, even when the number of votes are exactly known beforehand, for every $1-\delta > \frac{3\eps}{2}$.
\end{corollary}

For the $k$-veto voting rule, we prove below, again by reducing from \textsc{Augmented-indexing}, a slightly weaker space complexity lower bound compared to the bounds of \MakeUppercase theorem\nobreakspace \ref {thm:eps_eps}.

\begin{restatable}{theorem}{ThmVetoLB}\shortversion{[$\star$]}\label{thm:veto_lb}
 Suppose the number of candidates $m$ is at least $\frac{1}{\eps}$. Any one pass \textsc{$(\eps, \delta)$--Winner Determination} algorithm for the $k$-veto voting rule for $k=O(m^\lambda)$, for every $\lambda\in[0,1)$, must use $\Omega(\frac{1}{\eps^\mu}\log \frac{1}{\eps})$, for every constant $\mu <1$, bits of memory, even when the number of votes are exactly known beforehand, for every $1-\delta > \frac{3\eps}{2}$. 
\end{restatable}

\longversion{
\begin{proof}
 We prove the result for \textsc{$(\frac{\eps}{5}, \delta)$--Winner Determination} problem. Consider the \textsc{Augmented-indexing}$_{\frac{1}{\eps^{1-\mu}}, \frac{1}{\eps^\mu}}$ problem where the first player Alice is given a string $x\in [\frac{1}{\eps^{1-\mu}}]^\frac{1}{\eps^\mu}$, while the second player Bob is given an integer $i\in [\frac{1}{\eps^\mu}]$ and $x_j$ for every $j < i$. The candidate set of our election is $[\frac{1}{\eps^{1-\mu}}]\times[\frac{1}{\eps^\mu}]$. The votes would be such that the only $\frac{\eps}{5}$--winner will be the candidate $(x_i, i)$, thereby proving the result. Thus Bob can output $x_i$ correctly whenever our \textsc{$(\eps, \delta)$--Winner Determination} algorithm outputs correctly. Alice generates a stream of $\frac{n}{2}$ votes (assume $n$ to be sufficiently large) in such a way that for every $a,b\in \{(x_j,j) : j\in \frac{1}{\eps^\mu}\}$ and $x,y\in [\frac{1}{\eps^{1-\mu}}]\times[\frac{1}{\eps^\mu}] \setminus \{(x_j,j) : j\in \frac{1}{\eps^\mu}\}$, we have $s(a)-s(x)\ge \frac{\eps n}{2}$, $s(b)-1\le s(a)\le s(b)+1$, and $s(y)-1\le s(x)\le s(y)+1$, where $s(\cdot)$ is the number of vetoes that a candidate receives (which is always negative or zero). This is possible since $k=O(m^\lambda)$ for $\lambda\in[0,1)$. Alice now sends the memory content of the algorithm. Bob resumes the run of the algorithm by generating another stream of $\frac{n}{2}$ votes in such a way that for every $a^\prime,b^\prime\in \{(z,i) : z\in \frac{1}{\eps^{1-\mu}}\}$ and $x^\prime,y^\prime\in [\frac{1}{\eps^{1-\mu}}]\times[\frac{1}{\eps^\mu}] \setminus \{(z,i) : z\in \frac{1}{\eps^{1-\mu}}\}$, we have $s(a^\prime)-s(x^\prime)\ge \frac{\eps n}{2}$, $s(b^\prime)-1\le s(a^\prime)\le s(b^\prime)+1$, and $s(y^\prime)-1\le s(x^\prime)\le s(y^\prime)+1$. Now the score of $(x_i,i)$ is more than the score of every other candidate by at least $\frac{\eps n}{2}$. Hence, the candidate $(x_i, i)$ is the unique $\frac{\eps}{5}$--winner.
\end{proof}
}

For the $k$-approval voting rule, we provide a stronger space complexity lower bound of $\Omega(\frac{k}{\eps}\log\frac{1}{\eps})$, when the number of candidates $m$ is at least $\frac{k}{\eps^2}$, by reducing from \textsc{Augmented-indexing}$_{\frac{1}{\eps},\frac{k}{\eps}}$.

\begin{restatable}{theorem}{ThmKappLB}\shortversion{[$\star$]}\label{thm:kapp_lb}
 Assume that the number of candidates $m$ is at least $\frac{k}{\eps^2}$. Then any one pass \textsc{$(\eps, \delta)$--Winner Determination} algorithm for the $k$-approval voting rule must use $\Omega(\frac{k}{\eps}\log\frac{1}{\eps})$ bits of memory. 
\end{restatable}

\longversion{
\begin{proof}
 We prove the result for \textsc{$(\frac{\eps}{5}, \delta)$--Winner Determination} problem. Consider the \textsc{Augmented-indexing}$_{\frac{1}{\eps},\frac{k}{\eps}}$ problem where Alice is given $(x_1, \cdots, x_{\frac{k}{\eps}})\in[\frac{1}{\eps}]^\frac{k}{\eps}$ and Bob is given $(x_1, \cdots, x_{i-1})$. We will create a $k$-approval election in such a way that the $\frac{\eps}{5}$-winner will reveal $x_i$ to Bob. The candidate set of our election is $[\frac{1}{\eps}]\times[\frac{k}{\eps}]$. For every $j\in[k]$, Alice generates $\frac{\eps n}{2}$ many votes approving candidates in $\{(x_{k(j-1)+1},k(j-1)+1), (x_{k(j-1)+2},k(j-1)+2), \cdots, (x_{kj},kj)\}$. Alice now sends the memory content to Bob. Let $\mathcal{X} = \{ (j,i) : j\in[\frac{1}{\eps}] \}$. If $k\le\frac{1}{\eps}$ then, Bob generates $\frac{n}{2}$ votes in such a way that every candidate in $\mathcal{X}$ gets at least $\frac{k\eps n}{2}$ many approvals and the candidates in $[\frac{1}{\eps}]\times[\frac{k}{\eps}] \setminus \mathcal{X}$ does not get any approval from the votes that Bob generates. Now, the $k$-approval score of the candidate $(x_i,i)$ is at least $(k+1)\frac{\eps n}{2}$, whereas every other candidate gets at most $\frac{k \eps n}{2}$ many approvals. Hence, $(x_i,i)$ is the unique $\frac{\eps}{5}$-winner. If $k > \frac{1}{\eps}$ then, Bob generates $\frac{n}{2}$ votes in such a way that every candidate in $\mathcal{X}$ gets $\frac{n}{2}$ many approvals and every candidate in $[\frac{1}{\eps}]\times[\frac{k}{\eps}] \setminus \mathcal{X}$ gets at most $(k-\frac{1}{\eps})\frac{n}{2}\frac{1}{k/\eps^2 - 1/\eps}\le \frac{n}{2}\eps^2$ many approvals from the votes that Bob generates. Here again the $k$-approval score of the candidate $(x_i,i)$ is at least $(1+\eps)\frac{n}{2}$, where as the $k$-approval score of every other candidate is at most $\frac{\eps n}{2}$. Hence, $(x_i,i)$ is the unique $\frac{\eps}{5}$-winner.
\end{proof}
}

For the generalized plurality voting rule, we provide a $\Omega(\frac{1}{\sqrt{\eps}}\log m)$ space complexity lower bound, again by reducing from \textsc{Augmented-indexing}$_{m,\frac{1}{\sqrt{\eps}}}$. This bound is better than the lower bound of \MakeUppercase theorem\nobreakspace \ref {thm:eps_eps} when $m$ is exponentially larger compared to $\frac{1}{\eps}$.

\begin{restatable}{theorem}{ThmEpsLogm}\shortversion{[$\star$]}\label{thm:eps_logm}
 Suppose the number of candidates $m$ is at least $\frac{1}{\sqrt{\eps}}$. Any one pass \textsc{$(\eps, \delta)$--Winner Determination} algorithm for the generalized plurality rule must use $\Omega(\frac{1}{\sqrt{\eps}}\log m)$ bits of memory, for every $1-\delta > \frac{3\eps}{2}$. 
\end{restatable}

\longversion{
\begin{proof}
 We prove the result for \textsc{$(\frac{\eps}{5}, \delta)$--Winner Determination} problem. Consider the \textsc{Augmented-indexing}$_{m,\frac{1}{\sqrt{\eps}}}$ problem where Alice is given a string $x=(x_1, \cdots, x_{\frac{1}{\sqrt{\eps}}})\in [m]^{\frac{1}{\sqrt{\eps}}}$ and Bob is given an integer $i\in[\frac{1}{\sqrt{\eps}}]$ and $(x_1, \cdots, x_{i-1})$. The candidate set of our election is $[m]\times[\frac{1}{\sqrt{\eps}}]$. The votes would be such that the only $\frac{\eps}{5}$--winner will be the candidate $(x_i, i)$, thereby proving the result. Thus Bob can output $x_i$ correctly whenever our \textsc{$(\frac{\eps}{5}, \delta)$--Winner Determination} algorithm outputs correctly. Alice generates $(\frac{1}{\sqrt{\eps}}-j)\eps n$ many approvals for candidate $(x_j,j)$, for every $j<\frac{1}{\sqrt{\eps}}$. Alice now sends the memory content of the algorithm. Bob resumes the run of the algorithm by generating $(\frac{1}{\sqrt{\eps}}-j)\eps n$ many approvals for candidate $(x_j,j)$, for every $j<i$. Notice that, the only $\frac{\eps}{5}$-winner is the candidate $(x_i, i)$. Now the space complexity lower bound follows from \MakeUppercase lemma\nobreakspace \ref {lem:aug}.
\end{proof}
}

The space complexity lower bound in \MakeUppercase theorem\nobreakspace \ref {thm:eps_eps} for the plurality voting rule matches with the upper bound of \MakeUppercase theorem\nobreakspace \ref {thm:kapp}, when $\frac{1}{\eps} \le m \le \frac{1}{\eps^{O(1)}}$. For the case when $m\le \frac{1}{\eps}$, we now show a matching space complexity lower bound for the plurality voting rule.\longversion{ We prove this result by exhibiting a reduction from the \textsc{Max-sum}$_{\frac{1}{\eps},m}$ problem.}

\begin{theorem}\label{thm:mlog_eps}
 Assume that the number of candidates $m$ is at most $\frac{1}{\eps}$. Then any one pass \textsc{$(\eps, \delta)$--Winner Determination} algorithm for the plurality, generalized plurality, approval, $k$-approval for $k=O(m^\lambda)$, for any $\lambda\in[0,1)$, maximin, Copeland, Bucklin, plurality with run off voting rules must use $\Omega(m\log\frac{1}{\eps})$ bits of memory.
\end{theorem}

\shortversion{\vspace{-15pt}}\begin{proof}
 First, let us prove the result for the plurality voting rule. Suppose we have a one pass \textsc{$(\eps, \delta)$--Winner Determination} algorithm for the plurality election which uses $s(n,\eps)$ bits of memory. Consider the communication problem \textsc{Max-sum}$_{\frac{1}{\eps},m}$. Let the inputs to Alice and Bob in the \textsc{Max-sum}$_{\frac{1}{\eps},m}$ instance be $x=(x_1, x_2, \cdots, x_m)\in [\frac{1}{\eps}]^m$ and $y=(y_1, y_2, \cdots, y_m)\in [\frac{1}{\eps}]^m$ respectively. The candidate set of the election is $[m]$. Alice generates $x_i$ many plurality vote for the candidate $i$, for every $i\in[m]$. Alice now sends the memory content of the algorithm to Bob. Bob resumes the run of the algorithm by generating $y_i$ many plurality votes for the candidate $i$, for every $i\in[m]$. Suppose $i= \argmax_{j\in[m]}\{x_j+y_j\}$ (recall from \MakeUppercase definition\nobreakspace \ref {def:maxsum} that there exist unique element $i$ that maximizes $x_i+y_i$) and $\ell \ne \argmax_{j\in[m]}\{x_j+y_j\}$. Then we have the following: 
 \[\frac{(x_i+y_i)-(x_\ell +y_\ell)}{\sum_{j\in[m]}(x_j+y_j)} \ge \frac{\eps}{2m} \ge \frac{\eps^2}{2}\] 
 The first inequality follows from the fact that $(x_i+y_i)-(x_\ell +y_\ell)\ge 1$ and $\sum_{j\in[m]}x_j+y_j\le \frac{2m}{\eps}$. The second inequality follows from the assumption that $m\le \frac{1}{\eps}$. Hence, whenever the \textsc{$(\frac{\eps^2}{5}, \delta)$--Winner Determination} algorithm outputs an $\frac{\eps^2}{5}$-winner, Bob also outputs correctly in the \textsc{Max-sum}$_{\frac{1}{\eps},m}$ problem instance. 
 
 For the other voting rules, the idea is the same as above: we will generate votes in such a way that ensures that the candidate $i$ wins if $i= \argmax_{j\in[m]}\{x_j+y_j\}$ by a margin of at least one. Below, we only specify the votes to be generated for other voting rules.
 
 \begin{itemize}[topsep=2pt,leftmargin=15pt,itemsep=0pt]
  \item {\bf Generalized plurality, approval:} Follows immediately from the fact that every plurality election is a valid generalized plurality and approval election too.
  
  \item {\bf $k$-approval for $k=O(m^\lambda)$, for any $\lambda\in[0,1)$:} Alice (respectively Bob) generates $x_i$ (respectively $y_i$) many votes such that candidate $i$ gets $x_i$ many approvals and the rest $(k-1)x_i$ many approvals are equally distributed among other $m-1$ candidates.
  
  \item {\bf Borda, maximin, Copeland, Bucklin, plurality with run off:} Alice (respectively Bob) generates $x_i$ (respectively $y_i$) many votes of the form $i \succ \overrightarrow{{\mathcal{C}_{-i}}}$ and another $x_i$ (respectively $y_i$) many votes of the form $i \succ \overleftarrow{{\mathcal{C}_{-i}}}$, where $\overrightarrow{{\mathcal{C}_{-i}}}$ is an arbitrary but fixed order of the candidates in $\mathcal{C}\setminus\{i\}$ and $\overleftarrow{{\mathcal{C}_{-i}}}$ is the reverse order of $\overrightarrow{{\mathcal{C}_{-i}}}$.\qedhere
 \end{itemize}
\end{proof}

Now we show space complexity lower bounds that depend on the number of votes $n$. The result below is obtained by reducing from the \textsc{Greater-than}$_{n}$ problem. The lower bound is tight in the number of votes $n$.

\begin{restatable}{theorem}{ThmLogLogn}\shortversion{[$\star$]}\label{thm:loglogn}
 Any one pass \textsc{$(\eps, \delta)$--Winner Determination} algorithm for the plurality voting rule must use $\Omega(\log \log n)$ memory bits, even if the number of candidates is only $2$, for every $\delta < \frac{1}{4}$.  The same applies for generalized plurality, scoring rules, maximin, Copeland, Bucklin, and plurality with run off voting rules.  
\end{restatable}

\longversion{
\begin{proof}
 Suppose we have a one pass \textsc{$(\eps, \delta)$--Winner Determination} algorithm for the plurality election which uses $s(n)$ bits of space. Using this algorithm, we will show a communication protocol for the \textsc{Greater-than}$_{n}$ problem whose communication coplexity is $s(2^n)$ thereby proving the statement. The candidate set is $\{0,1\}$. Alice generates a stream of $2^x$ many plurality votes for the candidate $1$. Alice now sends the memory content of the algorithm. Bob resumes the run of the algorithm by generating a stream of $2^y$ many plurality votes for the candidate $0$. If $x>y$ then the candidate $1$ is the only $\eps$-winner; whereas if $x<y$ then the candidate $0$ is the only $\eps$-winner.

  For elections with two candidates, generalized plurality, scoring rules, maximin, Copeland, Bucklin, and plurality with run off voting rules are same as the plurality voting rule.
\end{proof}
}

\section{Conclusions and Future Work}

In this work, we studied the space complexity for determining approximate winners in the setting where votes are inserted continually into a data stream. We showed that allowing randomization and approximation indeed allows for much more space-efficient algorithms. Moreover, our bounds are tight in certain parameter ranges.

The most immediate open question is to close the gaps between the upper and lower bounds. In particular, even for plurality, the dependence on $m$ and $\eps$ is not tight when $m$ is large. Also, for the other voting rules, are there more sophisticated algorithms which improve our upper bounds? In a different vein, it may be interesting to implement these streaming algorithms for use in practice (say, for participatory democracy experiments or for online social networks) and investigate how they perform. Finally, instead of having the algorithm be passive, could we improve performance by having the algorithm actively query the voters as they appear in the stream?

\paragraph*{Acknowledgement:} We thank David Woodruff for helpful conversations about the heavy hitters problem.

\bibliographystyle{apalike}
\bibliography{sketch}

\newpage
\shortversion{

\section*{Appendix}

We now present the missing proofs.

\ObsSamplingUB*

\begin{proof}
 First let us assume, for simplicity, that $n$ is a power of $2$. We toss a fair coin $\log_2 n$ many times and choose the item, say $x$, only if the coin comes head all the times. Hence the probability that the item $x$ gets chosen is $\frac{1}{n}$. We need $O(\log\log n)$ space to toss the fair coin $\log_2 n$ times (to keep track of the number of times we have tossed the coin so far). If $n$ is not a power of $2$ then, toss the fair coin $\lceil \log_2 n \rceil$ many times and we choose the item $x$ only if the coin comes head in all the tosses conditioned on some event $E$. The event $E$ contains exactly $n$ outcomes including the all heads outcome.
\end{proof}

\PropSamplingLB*

\begin{proof}
 The algorithm tosses the fair coin some number of times (the number of times it tosses the coin may also depend on the outcome of the previous tosses) and finally picks an item from the set. Consider a run $\mathcal{R}$ of the algorithm where it chooses the item, say $x$, with {\em smallest number of coin tosses}; say it tosses the coin $t$ many times in this run $\mathcal{R}$. This means that in any other run of the algorithm where the item $x$ is chosen, the algorithm must toss the coin at least $t$ number of times. Let the outcome of the coin tosses in $\mathcal{R}$ be $r_1, \cdots, r_t$. Let $s_i$ be the memory content of the algorithm immediately after it tosses the coin $i^{th}$ time, for $i\in [t]$, in the run $\mathcal{R}$. First notice that if $t < \log_2 n$, then the probability with which the item $x$ is chosen is more than $\frac{1}{n}$, which would be a contradiction. Hence, $t \ge \log_2 n$. Now we claim that all the $s_i$'s must be different. Indeed otherwise, let us assume $s_i = s_j$ for some $i<j$. Then the algorithm chooses the item $x$ after tossing the coin $t-(j-i)$ (which is strictly less than $t$) many times when the outcome of the coin tosses are $r_1, \cdots, r_i, r_{j+1}, \cdots, r_t$. This contradicts the assumption that the run $\mathcal{R}$ we started with chooses the item $x$ with smallest number of coin tosses.
\end{proof}

\ThmMisra*

\begin{proof}
 The $O\left(\frac{1}{\eps}\left(\log m + \log n \right)\right)$ space algorithm is due to~\citep{misra82}. On the other hand, notice that with space $O\left(m\log n \right)$, we can exactly count the frequency of every element, {\em even in the turnstile model of stream}, by simply keeping an array of length $m$ (indexed by ids of the elements from the universe) each entry of which is capable of storing integers up to $n$.
\end{proof}
  
\KappUbThm*

\begin{proof}
 Let us first consider the case of the $k$-approval voting rule. We pick the current vote in the stream with probability $p$ (the value of $p$ will be decided later) independent of other votes. Suppose we sample $\ell$ many votes; let $\mathcal{S} = \{ v_i : i\in[\ell] \}$ be the set of votes sampled. From the set of sampled votes $\mathcal{S}$, we generate a stream $\mathcal{T}$ over the universe $\mathcal{C}$ as follows. For $i\in [\ell]$, let the vote $v_i$ be $c_1\succ c_2\succ \cdots \succ c_m$. From the vote $v_i$, we add $k$ candidates $c_1, \cdots, c_k$ in the stream $\mathcal{T}$. We know that there is a $\ell = O(\frac{\log(k+1)}{\eps^2} \log \frac{1}{\delta})$ (and thus a corresponding $p=\Omega(\frac{1}{n})$) which ensures that for every candidate $x\in \mathcal{C}$, $|\frac{s(x)}{n} - \frac{\hat{s}(x)}{\ell}| < \frac{\eps}{3}$ with probability at least $1 - \frac{\delta}{2}$~\citep{deysampling}, where $s(\cdot)$ and $\hat{s}(\cdot)$ are the scores of the candidates in the input stream of votes and in $\mathcal{S}$ respectively. Now we count $\hat{s}(x)$ for every candidate $x\in \mathcal{C}$ within an additive approximation of $\frac{\eps \ell}{3}$ and the result follows from \MakeUppercase theorem\nobreakspace \ref {thm:misra} (notice that the length of the stream $\mathcal{T}$ is $k\ell$). 
 
 For the $k$-veto voting rule, we approximately calculate the number of vetoes that every candidate gets using the same technique as above. However, for the $k$-veto voting rule, the corresponding bound for $\ell$ is $O(\frac{\log(m-k+1)}{\eps^2} \log \frac{1}{\delta})$ which implies the result.
\end{proof}

\ThmGenPluUB*

\begin{proof}
 We sample $\ell=O(\frac{1}{\eps^2}\log\frac{1}{\delta})$ many votes uniformly at random from the input stream of votes using the technique used in the proof of \MakeUppercase theorem\nobreakspace \ref {thm:kapp}. For every candidate, we count both the number of approvals and disapprovals that it gets within an additive approximation of $\frac{\eps\ell}{10}$ which is enough to get an $\eps$-winner. Now the space complexity follows form \MakeUppercase theorem\nobreakspace \ref {thm:misra}.
\end{proof}

\ThmApprovalUB*

\begin{proof}
 We sample $\ell$ many votes using the algorithm described in \MakeUppercase observation\nobreakspace \ref {lem:sampling_ub} and technique described in the proof of \MakeUppercase theorem\nobreakspace \ref {thm:scr}. The total number of approvals in those sampled votes is at most $m\ell$ and we estimate the number of approvals that every candidate receives within an additive approximation of $\frac{\eps\ell}{2}$. The result now follows from the upper bound on $\ell$~\citep{deysampling} and \MakeUppercase theorem\nobreakspace \ref {thm:misra}.
\end{proof}

\ThmStoreUB*

\begin{proof}
 We sample $\ell$ many votes from the input stream of votes uniformly at random and simply store all of them. Notice that we can store a vote using space $O(m\log m)$. The result now follows from the upper bound on $\ell$~\citep{deysampling} and \MakeUppercase observation\nobreakspace \ref {lem:sampling_ub}.
\end{proof}

\CorNUnknown*

\begin{proof}
 We use reservoir sampling with approximate counting from \MakeUppercase theorem\nobreakspace \ref {thm:sampling-unbd}. The resulting stream that we generate have both positive and negative updates (since in reservoir sampling, we sometimes replace an item we previously sampled). Now we approximately estimate the frequency of every item in the generated stream using \MakeUppercase theorem\nobreakspace \ref {thm:count-min}.
\end{proof}

\LemMaxSumLB*

\begin{proof}
We reduce the \textsc{Augmented-indexing}$_{2,t\log m}$ problem to \textsc{Max-sum}$_{8m,t+1}$ problem thereby proving the result. Let the inputs to Alice and Bob in the \textsc{Augmented-indexing}$_{2,t\log m}$ instance be $(a_1, a_2, \cdots, a_{t\log m})\in \{0,1\}^{t\log m}$ and $(a_1, \cdots, a_{i-1})$ respectively. The idea is to construct a corresponding instance of the \textsc{Max-sum}$_{8m,t+1}$ problem that outputs $t+1$ if and only if $a_i=0$. We achieve this as follows. Alice starts execution of the \textsc{Max-sum}$_{8m,t+1}$ protocol using the vector $x=(x_1, x_2, \cdots, x_{t+1})\in [8m]^{t+1}$ which is defined as follows: the binary representation of $x_j$ is $\left(0,0,a_{\left(j-1\right)\log m + 1}, a_{\left(j-1\right)\log m + 2}, a_{\left(j-1\right)\log m + 3}, \cdots, a_{j\log m}, 0\right)_2$, for every $j\in [t]$, and $x_{t+1}$ is $0$. Bob participates in the \textsc{Max-sum}$_{8m,t+1}$ protocol with the vector $y=(y_1, y_2, \cdots, y_{t+1})\in [8m]^{t+1}$ which is defined as follows. Let us define $\lambda = \lceil \frac{i}{\log m} \rceil$. We define $y_j=0$, for every $j\notin \{\lambda, t+1\}$. The binary representation of $y_\lambda$ is $(1,0, a_{(\lambda-1)\log m+1}, a_{(\lambda-1)\log m+2}, \cdots, a_{i-1}, 1, 0, 0, \cdots, 0, 0, 1)_2$. Let us define an integer $T$ whose binary representation is $(0,0, a_{(\lambda-1)\log m+1}, a_{(\lambda-1)\log m+2}, \cdots, a_{i-1}, 0, 1, 1, \cdots, 1)_2$. We define $y_{t+1}$ to be $T+y_\lambda$. First notice that the output of the \textsc{Max-sum}$_{8m,t+1}$ instance is either $\lambda$ or $t+1$, by the construction of $y$.  Now observe that if $a_i=1$ then, $x_\lambda>T$ and thus the output of the \textsc{Max-sum}$_{8m,t+1}$ instance should be $\lambda$. On the other hand, if $a_i=0$ then, $x_\lambda<T$ and thus the output of the \textsc{Max-sum}$_{8m,t+1}$ instance should be $t+1$.
\end{proof}

\LemGT*

\begin{proof}
 We reduce the \textsc{Augmented-indexing}$_{2,\lceil\log n\rceil + 1}$ problem to the \textsc{Greater-than}$_{n}$ problem thereby proving the result. Alice runs the \textsc{Greater-than}$_{n}$ protocol with its input number whose representation in binary is $a=(x_1x_2\cdots x_{\lceil\log n\rceil}1)_2$. Bob participates in the \textsc{Greater-than}$_{n}$ protocol with its input number whose representation in binary is $b=(x_1x_2\cdots x_{i-1}1\underbrace{0 \cdots 0}_{(\lceil\log n\rceil-i+1)~ 0's})_2$. Now $x_i=1$ if and only if $a>b.$
\end{proof}

\ThmExactLB*

\begin{proof}
 We prove the result for \textsc{$(0, \delta)$--Winner Determination} problem for the plurality election. This gives the result for the generalized plurality election since every plurality election is also a generalized plurality election. Consider the \textsc{Max-sum}$_{n,m}$ problem where Alice is given a string $x=(x_1, \cdots, x_m)\in[n]^m$ and Bob is given another string $y=(y_1, \cdots, y_m)\in[n]^m$. The candidate set of our election is $[m]$. The votes would be such that the only winner will be the candidate $i$ such that $i\in \argmax_{j\in[m]}\{x_j+y_j\}$. Moreover, the winner would be known to Bob, thereby proving the result. Thus Bob can output $x_i$ correctly whenever our \textsc{$(0, \delta)$--Winner Determination} algorithm outputs correctly. Alice generates $x_j$ many plurality votes for the candidate $j$, for every $j\in[m]$. Alice now sends the memory content to Bob. Bob resumes the run of the algorithm by generating $y_j$ many plurality votes for the candidate $j$, for every $j\in[m]$. The plurality score of candidate $j$ is $(x_j+y_j)$ and thus the plurality winner will be a candidate $i$ such that $i\in \argmax_{j\in[m]}\{x_j+y_j\}$. Notice that the total number of votes is at most $2mn$. The result now follows from \MakeUppercase lemma\nobreakspace \ref {mel:maxsum}.
\end{proof}

\ThmExactLBmn*
 
\begin{proof}
  Suppose we have a one pass \textsc{$(0, \delta)$--Winner Determination} algorithm for the plurality election that uses $s$ bits of memory. We will demonstrate a one-way three party protocol to compute \textsc{Disj}$_{m,3}^{promise}$ function using $2s$ bits of communication thus proving the result. We have the candidate set $[m+1]$. The protocol is as follows.
  
  Player $1$ starts running the one pass \textsc{$(0, \delta)$--Winner Determination} algorithm on the input $X_1\cup \{m+1\}$. Once player $1$ is done reading all its input, it sends its memory content to player $2$. This needs at most $s$ bits of communication. Player $2$ resumes the run of the algorithm with input $X_2\cup \{m+1\}$ and sends its memory content to player $3$. Again this needs at most $s$ bits of communication. Player $3$ resumes the run of the algorithm on input $X_3$ and output $1$ if and only if the winner is $m+1$ and $0$ else. Notice that, if the $X_i \cap X_j = \emptyset$ for every $i\ne j$ then, the only winner of the votes $(X_1, m+1, X_2, m+1, X_3)$ is the candidate $m+1$ with a plurality score of two. On the other hand, if there exist an element $y\in [m]$ such that $y\in X_i$ for every $i\in [t]$ and $(X_i\setminus\{y\}) \cap (X_j\setminus\{y\}) = \emptyset$ for every $i\ne j$ then, the only winner of the votes $(X_1, m+1, X_2, m+1, X_3)$ is the candidate $y$ with a plurality score of three.
  
  The number of candidates in the election above is $m+1$ and the number of votes $n$ is $|X_1|+|X_2|+|X_3|+2(m+1) = \Theta(m).$ This gives a space complexity lower bound of $\Omega(\min\{m,n\}).$
\end{proof}

\ThmLogn*

\begin{proof}
 For the sake of contradiction, we assume that the number of possible memory contents of the algorithm is $o(n)$, since otherwise the algorithm uses $\Omega(\log n)$ space and we have nothing to prove. Our candidate set is $\{0,1\}$.  We will generate two vote streams, say $R_1$ and $R_2$, in such a way that the final state of the algorithm would be same; however $\eps$--winner would be different for the two streams thus providing the contradiction we are looking for. 
 
 Let $s_0$ be the starting state of the algorithm. Consider the stream of votes for $1$ and let the algorithm repeats its state for the first time after reading $i$ many $1$ votes. Let the state of the algorithm after reading $i^{th}$ $1$ vote be same as the state the algorithm was after it read $j^{th}$ $1$ vote. Let us call $\mu = i-j$. Clearly $\mu = o(n)$. Then there exist $\delta_1, \delta_2 = o(n)$ such that the state the algorithm will be after reading $\frac{n}{4}-\delta_1$ many votes for $1$ is same as the state it will be after reading $\frac{3n}{4}+\delta_2$ many votes for $1$. Let $R_1$ be the stream of $\frac{n}{4}-\delta_1$ many votes for $1$ followed by $\frac{n}{2}$ many votes for $0$. Let $R_2$ be the stream of $\frac{3n}{4}+\delta_2$ many votes for $1$ followed by $\frac{n}{2}$ many votes for $0$. By construction the output of the algorithm is same for both the streams $R_1$ and $R_2$. However, candidate $1$ is only $\eps$-winner in $R_1$ and candidate $0$ is only $\eps$-winner in $R_2$.

For elections with two candidates, scoring rules, maximin, Copeland, Bucklin, and plurality with run off voting rules are same as the plurality voting rule.
\end{proof}

\ThmVetoLB*

\begin{proof}
 We prove the result for \textsc{$(\frac{\eps}{5}, \delta)$--Winner Determination} problem. Consider the \textsc{Augmented-indexing}$_{\frac{1}{\eps^{1-\mu}}, \frac{1}{\eps^\mu}}$ problem where the first player Alice is given a string $x\in [\frac{1}{\eps^{1-\mu}}]^\frac{1}{\eps^\mu}$, while the second player Bob is given an integer $i\in [\frac{1}{\eps^\mu}]$ and $x_j$ for every $j < i$. The candidate set of our election is $[\frac{1}{\eps^{1-\mu}}]\times[\frac{1}{\eps^\mu}]$. The votes would be such that the only $\frac{\eps}{5}$--winner will be the candidate $(x_i, i)$, thereby proving the result. Thus Bob can output $x_i$ correctly whenever our \textsc{$(\eps, \delta)$--Winner Determination} algorithm outputs correctly. Alice generates a stream of $\frac{n}{2}$ votes (assume $n$ to be sufficiently large) in such a way that for every $a,b\in \{(x_j,j) : j\in \frac{1}{\eps^\mu}\}$ and $x,y\in [\frac{1}{\eps^{1-\mu}}]\times[\frac{1}{\eps^\mu}] \setminus \{(x_j,j) : j\in \frac{1}{\eps^\mu}\}$, we have $s(a)-s(x)\ge \frac{\eps n}{2}$, $s(b)-1\le s(a)\le s(b)+1$, and $s(y)-1\le s(x)\le s(y)+1$, where $s(\cdot)$ is the number of vetoes that a candidate receives (which is always negative or zero). This is possible since $k=O(m^\lambda)$ for $\lambda\in[0,1)$. Alice now sends the memory content of the algorithm. Bob resumes the run of the algorithm by generating another stream of $\frac{n}{2}$ votes in such a way that for every $a^\prime,b^\prime\in \{(z,i) : z\in \frac{1}{\eps^{1-\mu}}\}$ and $x^\prime,y^\prime\in [\frac{1}{\eps^{1-\mu}}]\times[\frac{1}{\eps^\mu}] \setminus \{(z,i) : z\in \frac{1}{\eps^{1-\mu}}\}$, we have $s(a^\prime)-s(x^\prime)\ge \frac{\eps n}{2}$, $s(b^\prime)-1\le s(a^\prime)\le s(b^\prime)+1$, and $s(y^\prime)-1\le s(x^\prime)\le s(y^\prime)+1$. Now the score of $(x_i,i)$ is more than the score of every other candidate by at least $\frac{\eps n}{2}$. Hence, the candidate $(x_i, i)$ is the unique $\frac{\eps}{5}$--winner.
\end{proof}

\ThmKappLB*

\begin{proof}
 We prove the result for \textsc{$(\frac{\eps}{5}, \delta)$--Winner Determination} problem. Consider the \textsc{Augmented-indexing}$_{\frac{1}{\eps},\frac{k}{\eps}}$ problem where Alice is given $(x_1, \cdots, x_{\frac{k}{\eps}})\in[\frac{1}{\eps}]^\frac{k}{\eps}$ and Bob is given $(x_1, \cdots, x_{i-1})$. We will create a $k$-approval election in such a way that the $\frac{\eps}{5}$-winner will reveal $x_i$ to Bob. The candidate set of our election is $[\frac{1}{\eps}]\times[\frac{k}{\eps}]$. For every $j\in[k]$, Alice generates $\frac{\eps n}{2}$ many votes approving candidates in $\{(x_{k(j-1)+1},k(j-1)+1), (x_{k(j-1)+2},k(j-1)+2), \cdots, (x_{kj},kj)\}$. Alice now sends the memory content to Bob. Let $\mathcal{X} = \{ (j,i) : j\in[\frac{1}{\eps}] \}$. If $k\le\frac{1}{\eps}$ then, Bob generates $\frac{n}{2}$ votes in such a way that every candidate in $\mathcal{X}$ gets at least $\frac{k\eps n}{2}$ many approvals and the candidates in $[\frac{1}{\eps}]\times[\frac{k}{\eps}] \setminus \mathcal{X}$ does not get any approval from the votes that Bob generates. Now, the $k$-approval score of the candidate $(x_i,i)$ is at least $(k+1)\frac{\eps n}{2}$, whereas every other candidate gets at most $\frac{k \eps n}{2}$ many approvals. Hence, $(x_i,i)$ is the unique $\frac{\eps}{5}$-winner. If $k > \frac{1}{\eps}$ then, Bob generates $\frac{n}{2}$ votes in such a way that every candidate in $\mathcal{X}$ gets $\frac{n}{2}$ many approvals and every candidate in $[\frac{1}{\eps}]\times[\frac{k}{\eps}] \setminus \mathcal{X}$ gets at most $(k-\frac{1}{\eps})\frac{n}{2}\frac{1}{k/\eps^2 - 1/\eps}\le \frac{n}{2}\eps^2$ many approvals from the votes that Bob generates. Here again the $k$-approval score of the candidate $(x_i,i)$ is at least $(1+\eps)\frac{n}{2}$, where as the $k$-approval score of every other candidate is at most $\frac{\eps n}{2}$. Hence, $(x_i,i)$ is the unique $\frac{\eps}{5}$-winner.
\end{proof}

\ThmEpsLogm*

\begin{proof}
 We prove the result for \textsc{$(\frac{\eps}{5}, \delta)$--Winner Determination} problem. Consider the \textsc{Augmented-indexing}$_{m,\frac{1}{\sqrt{\eps}}}$ problem where Alice is given a string $x=(x_1, \cdots, x_{\frac{1}{\sqrt{\eps}}})\in [m]^{\frac{1}{\sqrt{\eps}}}$ and Bob is given an integer $i\in[\frac{1}{\sqrt{\eps}}]$ and $(x_1, \cdots, x_{i-1})$. The candidate set of our election is $[m]\times[\frac{1}{\sqrt{\eps}}]$. The votes would be such that the only $\frac{\eps}{5}$--winner will be the candidate $(x_i, i)$, thereby proving the result. Thus Bob can output $x_i$ correctly whenever our \textsc{$(\frac{\eps}{5}, \delta)$--Winner Determination} algorithm outputs correctly. Alice generates $(\frac{1}{\sqrt{\eps}}-j)\eps n$ many approvals for candidate $(x_j,j)$, for every $j<\frac{1}{\sqrt{\eps}}$. Alice now sends the memory content of the algorithm. Bob resumes the run of the algorithm by generating $(\frac{1}{\sqrt{\eps}}-j)\eps n$ many approvals for candidate $(x_j,j)$, for every $j<i$. Notice that, the only $\frac{\eps}{5}$-winner is the candidate $(x_i, i)$. Now the space complexity lower bound follows from \MakeUppercase lemma\nobreakspace \ref {lem:aug}.
\end{proof}

\ThmLogLogn*

\begin{proof}
 Suppose we have a one pass \textsc{$(\eps, \delta)$--Winner Determination} algorithm for the plurality election which uses $s(n)$ bits of space. Using this algorithm, we will show a communication protocol for the \textsc{Greater-than}$_{n}$ problem whose communication coplexity is $s(2^n)$ thereby proving the statement. The candidate set is $\{0,1\}$. Alice generates a stream of $2^x$ many plurality votes for the candidate $1$. Alice now sends the memory content of the algorithm. Bob resumes the run of the algorithm by generating a stream of $2^y$ many plurality votes for the candidate $0$. If $x>y$ then the candidate $1$ is the only $\eps$-winner; whereas if $x<y$ then the candidate $0$ is the only $\eps$-winner.

  For elections with two candidates, generalized plurality, scoring rules, maximin, Copeland, Bucklin, and plurality with run off voting rules are same as the plurality voting rule.
\end{proof}

}

\end{document}